%% file: main.tex
\documentclass[sigconf]{acmart}

\usepackage{amssymb}
\AtBeginDocument{%
}

\usepackage{graphicx}
\usepackage{subcaption}
\usepackage{amsmath,amsfonts,amssymb}
\usepackage{booktabs}
\usepackage{multirow}
\usepackage{hyperref}

\usepackage{mathtools,}

\newcommand{\heading}[1]{{\vspace*{6pt}\noindent\bf{#1 }}}
\newcommand{\new}[1]{{\color{black}#1}}

\newcommand{\twisha}[1]{{\color{black}#1}}

\copyrightyear{2025}
\acmYear{2025}
\setcopyright{cc}
\setcctype{by}
\acmConference[CCS '25]{Proceedings of the 2025 ACM SIGSAC Conference on Computer and Communications Security}{October 13--17, 2025}{Taipei, Taiwan}
\acmBooktitle{Proceedings of the 2025 ACM SIGSAC Conference on Computer and Communications Security (CCS '25), October 13--17, 2025, Taipei, Taiwan}\acmDOI{10.1145/3719027.3744837}
\acmISBN{979-8-4007-1525-9/2025/10}

\begin{document}
\title{One Video to Steal Them All: 3D-Printing IP Theft through Optical Side-Channels}


\author{Twisha Chattopadhyay}
\authornotemark[1]
\affiliation{%
  \institution{Georgia Institute of Technology}
  \city{Atlanta}
  \country{USA}}
\email{twishac@gatech.edu}

\author{Fabricio Ceschin}
\affiliation{%
  \institution{Todyl Inc.}
  \city{Atlanta}
  \country{USA}}
\email{fceschin@todyl.com}

\author{Marco E. Garza}
\affiliation{%
  \institution{The University of Texas at San Antonio}
  \city{San Antonio}
  \country{USA}}
\email{marco.garza@my.utsa.edu}

\author{Dymytriy Zyunkin}
\affiliation{%
  \institution{Georgia Institute of Technology}
  \city{Atlanta}
  \country{USA}}
\email{dzyunkin3@gatech.edu}

\author{Animesh Chhotaray}
\affiliation{%
  \institution{Georgia Institute of Technology}
  \city{Atlanta}
  \country{USA}}
\email{achhotaray3@gatech.edu}

\author{Aaron P. Stebner}
\affiliation{%
  \institution{Georgia Institute of Technology}
  \city{Atlanta}
  \country{USA}}
\email{aaron.stebner@gatech.edu}

\author{Saman Zonouz}
\affiliation{%
  \institution{Georgia Institute of Technology}
  \city{Atlanta}
  \country{USA}}
\email{saman.zonouz@gatech.edu}

\author{Raheem Beyah}
\affiliation{%
  \institution{Georgia Institute of Technology}
  \city{Atlanta}
  \country{USA}}
\email{rbeyah@ece.gatech.edu}

\renewcommand{\shortauthors}{Twisha et al.}

\input{Sources/Sections/abstract}   

\ccsdesc[500]{Security and privacy~Side-channel analysis and countermeasures}

\keywords{3D printing, reverse engineering, side-channel attack, security}

\maketitle

\setlength{\textfloatsep}{5pt}

\input{Sources/Sections/introduction}

\input{Sources/Sections/related-work-ac}

\input{Sources/Sections/background}

\input{Sources/Sections/methodology}

\input{Sources/Sections/evaluation}

\input{Sources/Sections/discussion}

\input{Sources/Sections/acknowledgement}

\bibliographystyle{ACM-Reference-Format}
\input{main.bbl}

\end{document}

%% file: Sources/Sections/abstract.tex
\begin{abstract}
\label{sec:Abs}
The 3D printing industry is rapidly growing and is increasingly adopted in various sectors, including manufacturing, healthcare, and defense. However, the operational setup often involves hazardous environments, necessitating remote monitoring through cameras and other sensors, which opens the door to cyber-based attacks. In this paper, we show that an adversary with access to video recordings of the 3D printing process can reverse-engineer the underlying 3D print instructions. Our model tracks the printer nozzle's movements during the printing process and maps the corresponding trajectory into G-code instructions. Further, it identifies the correct parameters such as feed rate and extrusion rate, which enable successful IP theft. To validate the success of IP theft, we design an equivalence checker that quantitatively compares two sets of 3D print instructions, evaluating their similarity in producing objects that are alike in shape, external appearance, and internal structure.
Our equivalence checker, unlike other simple distance-based metrics such as normalized mean square error, is rotationally as well as translationally invariant. This is necessary to capture shifts in the base/start position of the reverse-engineered instructions relative to the actual 3D print instructions that can happen due to different camera positions. 
Our model achieves an average accuracy of 90.87\% and generates 30.20\% fewer instructions compared to the current state-of-the-art methods that produce instructions that either lead to faulty or incorrect (in terms of differences in shape and internal structure) 3D prints. Additionally, we use our model to reverse-engineer the 3D print instructions from a video recording and print a fully-functional counterfeit object.


%


\end{abstract}

%% file: Sources/Sections/introduction.tex
\section{Introduction} 
\label{Section: Introduction}

The global additive-manufacturing (AM)  aka 3D printing industry was valued at around \$20 billion in 2023 \cite{MarketValue2024}. It is readily being adopted by different sectors considering its ability to produce complex geometries, reduce material waste, and shorten production cycles. For instance, in the healthcare sector, AM is used to print custom implants and prosthetics; the global 3D printing healthcare market is projected to reach \$3.7 billion by 2026 \cite{yu2023xcheck}. In the automotive sector, AM is used to manufacture low-cost lightweight components, prototypes, and custom parts, and enhance vehicle performance and fuel efficiency \cite{bmw2020additive,ford2023additive,gm2022additive}. Even governments (e.g. USA, France) are increasingly favoring the use of AM in the production of military equipment \cite{USAMAdoption,shop3duniverse2024military}, given the compact and speedy nature of 3D printers and 3D printing respectively.

In a high-valued 3D printing market, the 3D design model, typically described using an STL\footnote{STL stands for standard tesselation language and uses triangular meshes, typically represented as four triplets ---~ three vertices and a norm vector ---~ to describe the 3D model of an object.} file,  is the intellectual property (IP) of the designer. 
An IP owner customizes it with parameters that define the mechanical properties of the object that will be 3D printed. A design-compiler, commonly known as \textit{slicer}~\cite{ultimakercura} in 3D printing, is  typically used to convert the STL file with all the parameters, to form a sequence of geometric code~(G-code) instructions. These instructions will be interpreted by the 3D printer firmware~\cite{marlinfw} instruction-by-instruction to print the object \textit{additively}. For example, in filament-based 3D printing~\cite{williams2015additive}, an extruder heats and melts a plastic filament, which is then deposited additively layer by layer onto a print plate by the printer's nozzle; as each layer cools and solidifies, the printer precisely adds subsequent layers, gradually forming the final 3D object. The movement of the nozzle, as well as the information on when and by how much to extrude, is defined by a sequence of G-code instructions (stored in a G-code file).

Stealing the G-code IP, i.e., the G-code file, allows an adversary to produce counterfeit~\cite{kikuchi2018embedding,chen2019embedding,gultekin2019embedding} 3D printed objects.  Except for three G-code IP-theft attacks~\cite{do2016data,alkofahi2024mitm,sturm2017cyber} that assume an adversary compromised either the network used to transfer the G-code and printer configurations, or the CAD/slicer software \cite{sturm2017cyber}, all prior works aimed at reverse-engineering the G-code IP using only access to some side channels---~optical~\cite{liang2022hiding}, acoustic~\cite{mativo2018cyber,chhetri2017confidentiality,chhetri2017side,al2016acoustic,song2016smartphone}, power~\cite{gatlin2021encryption}, thermal~\cite{faruque2016forensics}. Since commercial 3D printing can produce hazardous emissions while melting plastic, metal, etc., these side channels are available via sensor/monitoring equipment that is used for in-situ \textit{remote} quality checks. For example, there are several solutions~\cite{gao2018watching,3DQue_optical,AstroPrint_optical,3DPrinterOS_optical} that allow operators to use cameras ---~an optical side channel---~ to monitor the 3D printing process.

\begin{figure*}[ht]
    \centering
    \includegraphics[width=\linewidth]{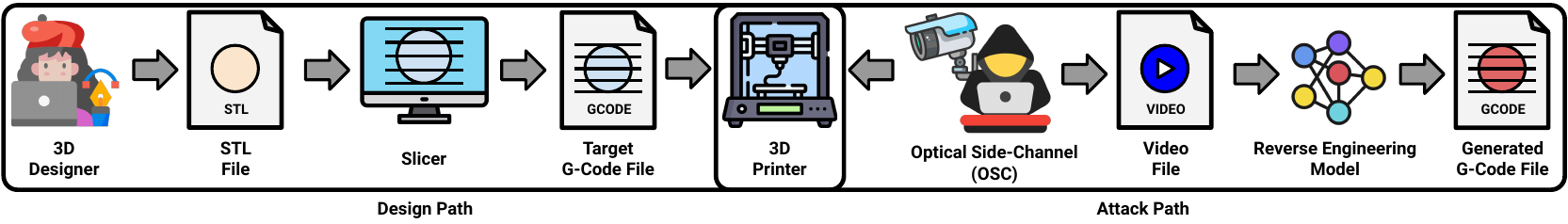}
    \caption{\textbf{Threat model.} A 3D designer creates a 3D object and slices it to generate a G-code that can be printed by the 3D printer (design path). The attacker can record the print process to obtain a video, which is used as input to a reverse engineering model that generates the corresponding G-code (attack path).}
    \label{fig:threat-model}
    \hrulefill
\end{figure*}

\textbf{Threat Model.} In this work, we assume that the adversary has exploited  vulnerabilities in IP-based cameras~\cite{liranzo2017security} to get the video recording of the printing process, and its goal is to reverse-engineer a \textit{3D printable} G-code IP  \textit{just} from the video recording. See \autoref{fig:threat-model} for a pictorial description of the threat model. 
\new{Note that we do not assume that the adversary has  planted any external device (e.g., smartphones~\cite{song2016smartphone}, hidden camera~\cite{liang2022hiding}) to gain access to the side channel. Rather, the camera is part of the 3D printing setup that is used by the operators to remotely monitor the 3D printing process, prevent material wastage, and perform post-mortem analysis in case of print failures.  Additionally, we assume that} the adversary has \textit{no} knowledge of the original G-code, i.e., they don't know the number of instructions the original \new{G-code} contains, the parameters that were set in order to create the G-code, the number of layers, or the ratio of instructions across layers. \new{Further}, the adversary has no knowledge regarding the position of the object being printed on the printing plate. \new{Lastly, we assume that the video feed from the camera contains the full 3D printer at all times; however, this does not give the adversary information about the coordinate reference frame set by the operator for the printing plate.}

No prior works have shown that such an attack is feasible in this threat model for realistically complex parts.  Moreover, \textit{no suitable metric exists} to compare the similarities between the original and the reverse-engineered G-codes. The need for such a metric is essential from both an attack and defense standpoint. From the attacker's perspective, such a metric helps determine whether the attacker could successfully steal the G-code IP. From a defense perspective, black-market G-codes can be compared against proprietary G-codes to detect IP theft. Note that the G-code IP/file contains a sequence of G-code instructions, where each G-code instruction typically describes a small part of the trajectory of the nozzle along with information about whether the nozzle's movements are accompanied by extrusion or not. Therefore, minimally, a reverse-engineered 3D printable G-code IP must not only describe the 3D path/trajectory traced by the nozzle while it is printing, it also has to encode how much the nozzle has to extrude during each small movement.

To the best of our knowledge, we create the first model that can reverse-engineer a \textit{3D printable} G-code IP from a video of the target object being 3D printed. The reverse-engineered G-code file describes the 3D trajectory of the nozzle, where the 3D trajectory is clearly separated into a set of 2D trajectories that define each layer of the object that is additively printed. 

Additionally, we use our model to classify whether each print instruction is an extruding instruction or not; for extruding instructions, we also determine the rate of extrusion by finding the distance traveled by the nozzle during the execution of an extruding print instruction. We note that, in a recent work (NDSS'22), Liang et al.~\cite{liang2022hiding} showed for the first time that a RESNET-50~\cite{krizhevsky2017imagenet} machine-learning model can be used to recover a 3D trajectory of the nozzle. However, we found that separating the 3D trajectory, which is recovered using Liang et al.'s RESNET-50 model, into distinct layers is very hard. This makes the process of converting the recovered 3D trajectory into a printable G-code infeasible in practice. Even if we somehow manage to achieve this, without extruding information, the 3D printer will most likely under or over-extrude every single time.

We also design a new \textit{G-code equivalence checker} that compares the similarity of two G-codes quantitatively. In other words, if two similar G-codes are fed to a 3D printer, we get two objects that are similar in terms of both the \textit{external shape}, as well as the \textit{internal design}. The G-code equivalence checker is an essential metric that is used to determine whether an IP theft attack was successful. Loosely, it translates to extracting the 3D trajectory of the nozzle from both G-codes and then finding the maximum overlap between the two trajectories. 
Standard distance metrics such as mean-squared error (MSE), or its normalized version (nMSE), were used by Liang et al.~\cite{liang2022hiding}.  Using nMSE during our experiments, a rotated version of an object was seen to be up to 45.82\% dissimilar to the G-code of its non-rotated counterpart. In other words, unless the attacker managed to reverse-engineer the counterfeit in the \textit{exact same position on the printing plate} as the original, the evaluation done by standard distance metrics would conclude the attack to have failed. This observation led us to create our G-code equivalence checker to be invariant to rotation and translation. To achieve this, we develop an \textit{oriented bounding polygon} (OBP) algorithm that is effectively a more generalized version of the well-known (in Computer Vision domain) oriented bounding box~\cite{9996428} algorithm.

In summary, we make the following contributions:
\vspace{-0.05in}
\begin{itemize}
     \item We build a 3D printing-specific G-code equivalence checker that is rotationally and translationally invariant. 
     \item We create a G-code-to-video dataset of 16 objects with multiple variations (e.g., by changing the camera angle) from their open-source 3D models. 
     \item We give the first solution that accurately  reverse-engineers a printable G-code IP given only the video recording of the print process. 
    \item Using our reverse-engineering model, we create a fully functional \textit{counterfeit} padlock key \new{and a gear system}.
   \end{itemize}

In the absence of an open-source dataset for training our ML model, we create a \textit{new} G-code-to-video dataset\footnote{We chose simple objects in the dataset because 3D printing can take hours to print even simple objects. The dataset can be found at \url{https://shorturl.at/v8fpY}} from scratch. The dataset consists of 16 different objects such as gears, keys, etc, whose 3D models are available on the open-source Thingiverse~\cite{thingiverse} platform. We slice the 3D models (as well as their rotated and translated versions) of these objects to get their G-codes, and then record (from two different camera angles) the 3D printing of the objects. Upon evaluating our G-code reverse-engineering solution on this dataset, we find that our solution recovers a G-code that is on average 90.87\% similar to the original G-code. \new{We also show that our reverse-engineering solution is (a) robust to change in camera angles by evaluating our solution on a separate G-code-to-video dataset where the videos were recorded from a different camera angle, and (b) robust to change in 3D printers by recovering the G-code instructions using videos of a Geetech printer, and 3D  printing the target object (counterfeit key) on an Ultimaker printer.}

We also create an additional dataset for checking the rotational and translational invariance of our G-code equivalence checker. Given the G-code of an object, we choose distance-offset values and then add the offset value to the (X,Y) coordinate in each G-code instruction to get a translated/shifted version of the G-code. For getting the rotated version, for a given degree offset, we extract the (X,Y) coordinates from the G-code instructions that define the trajectory of the nozzle of each layer, connect the points described by the (X,Y) coordinates to form an irregular polygon, find the centroid of the polygon and then rotate the vector connecting the centroid and each point by the degree-offset value. When compared with nMSE (used in related works), our curve checker, on average, assigns 99.76\%, 99.71\%, and 99.54\% similarity values to rotated, translated, and both rotated-and-translated variants of the same G-code, respectively; in comparison, average nMSE values on the same dataset are 74.65\%, 28.39\%  and 73.83\%. This implies that an nMSE-based checker does not work even when the reverse-engineered G-code produces an identical object but the base position is rotated/translated on the build plate.

%% file: Sources/Sections/related-work-ac.tex
\section{Related Works}
\label{sec:related-work}

\heading{Side-channel attacks.}  Liang et al. \cite{liang2022hiding} showed that using a ResNet-50 ML model, an adversary can recover the \new{three}-dimensional trajectory of the print head from a video recording. \new{Note that the adversarial goal in this case is different from our work as it does not require the model to produce \textit{3D-printable} G-code instructions. Recall that 3D printing is an additive process and the G-code instructions must describe both nozzle movement and extrusion timing.  The authors  \cite{liang2022hiding} also assume that an adversary can either plant a hidden camera (in the ceiling) in which case the camera is \textit{not fixed} or it can compromise the \textit{fixed} cameras used for surveillance or remote monitoring.
In our threat model, we assume the latter setup as it is more realistic. That means defenses such as noise injection to degrade the quality of the optical side-channel are not applicable against our reverse-engineering solution that uses an ML model along with a post-processor to recover valid 3D printing instructions that encode the trajectory \textit{and} the extrusion timing information of the nozzle.
}

There are other works \cite{al2016acoustic, chhetri2017confidentiality, chhetri2017side} that exploited \textit{acoustic} side-channels to reverse-engineer the trajectory as well as the extrusion rate.  Faruque et al. \cite{faruque2016forensics} and Gatlin et al. used \textit{thermal} and \textit{power} side-channels, respectively, to recover the trajectory of the print head. Some prior works \cite{gao2018watching}  used multiple side-channels (optical, acoustic, magnetic, acceleration) to recover 3D print parameters like infill patterns, printing speed, layer thickness, etc. Note that these works arguably consider a \textit{stronger} threat model as they assume that the adversary will be able to ``plant'' a device (e.g., phone, oscilloscope) to collect the side-channel information. Moreover, prior works used distance-based metrics (e.g., mean-squared error) to compare the predicted G-code/trajectory of the print head with the ground truth. Since distance-based metrics like Mean Squared Error (MSE) are \textit{not} rotationally or translationally invariant, it is hard to gauge the actual efficacy of the reverse-engineering tools built by prior works. Moreover, none of the prior works compared the overhead (in terms of the number of instructions) of the reverse-engineered G-code. In this work, we tackle all of these issues. 

\heading{ML-based object tracking.} Video object tracking combines object detection and tracking to follow the movement of objects across frames. In our threat model, one aspect of G-code reverse-engineering involves tracing the path (in the video) that the nozzle of a 3D printer took to print the 3D object, and mapping it to a path in the frame of reference of the build plate. However, tracking the movement of the nozzle of a 3D printer from a video presents unique challenges: high-fidelity tracking of small (in millimeter scale)  movements of a small object that current state-of-the-art object tracking algorithms (DeepSORT~\cite{wojke2017simple}, FairMOT~\cite{zhang2021fairmot}, and YOLO (You Only Look Once)~\cite{redmon2016you,bochkovskiy2020yolov4,li2022yolov6,chen2023yolo}) struggle to address.  For example,  YOLO/DeepSORT are designed to track multiple objects.  However, they fail to detect and track very small-scale movements accurately. Moreover, extensions of YOLO and DeepSORT (using SiamRPN~\cite{li2018high} and SiamMask~\cite{wang2019fast}) improve robustness by learning general appearance changes over time. These models, however, can be computationally expensive to build and often require extensive offline training for improved performance. Traditional algorithms like Mean-Shift~\cite{comaniciu2000real} and CamShift~\cite{bradski1998computer} use color histograms to track objects but struggle with scale and orientation changes. Optical flow techniques~\cite{horn1981determining} also analyze the motion of objects between consecutive frames for tracking~\cite{mahbub2011optical}. Despite their ability to capture fine-grained motion details, optical flow methods can be computationally intensive and may struggle with fast-moving objects and significant changes in object appearance. Real-time object detectors such as DETR~\cite{carion2020end} show promising performance but often fail to adapt to novel domains without pre-training on new datasets.

%% file: Sources/Sections/background.tex
\section{Background} 
\label{Section: Background}
\begin{figure}[h]
    \centering
    \includegraphics[width=1\linewidth]{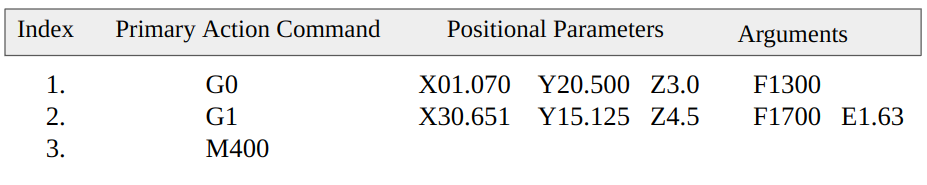}
    \caption{Examples of G-code Instructions}
    \hrulefill
    \label{gen}
\end{figure}

\heading{G-code instructions.} An object to be 3D printed begins with the creation of a 3D model that is typically described using STL~\cite{wong2012review}. The STL/design file is then passed through a slicer to generate a  G-code file that contains the sequence of G-code instructions for a 3D printer. While the design file is generic, a G-code contains all the parametric specifications needed to print the object. Since G-code is a coding language specific to 3D printers, it follows a fixed syntax as shown in \autoref{gen}. It can be seen that each instruction can be divided into three major components: Primary Action Command, Positional Parameters, and Arguments. 

The \textit{primary action commands} describe the type of movement the mechanical components of the 3D Printer is supposed to make from its current position to the destination position that is described using \textit{positional parameters}. The primary action commands can be used for various tasks. The \textbf{Gx} commands are used to describe the movement type (e.g., straight-line, curve) of the nozzle aka printhead; in this work, we consider linear movements (G0 and G1) as they are supported by all 3D printers. Other than print head movements, primary action commands can also be used to play a beep, raise the printing platform, etc. For example, the M300 \cite{marlin_m300} command (for a 3D printer that uses the Marlin firmware \cite{marlinfw}) plays a beep after a sequence of instructions has finished executing. In \autoref{gen}, we see examples of G0 and G1 being used in a G-code. The first instruction line tells the print nozzle to go to the 3D coordinate (1.07,10.5,3) without extruding. The second instruction line tells the print head to go to coordinates (30.651,15.1125,4.5) from the current position while extruding filament at a rate of 1.63mm/s. 

The \textit{arguments} dictate the mechanical actions of the filament during any and all movements of the print nozzle. The argument \textbf{E} gives the rate of extrusion while the nozzle is tracing a trajectory. The extrusion rate can be calculated deterministically for a given printer by using the equation $E = (4 h s l d_{n})d_{f}^2/\pi$, where $h$ is the layer height (0.3 mm in our slicer configuration), $s$ is the flow modifier (100\% in our printer recommended configuration), $d_{n}$ is the extruder nozzle diameter (0.4 in our printer), $d_{f}$ is the filament diameter (1.75 mm in our printer), and $l$ is the distance of the straight line (from point $(X_{n-1}, Y_{n-1})$ to point $(X_{n}, Y_{n})$). The argument 
\textbf{F} is used to specify the feed rate of the nozzle, i.e., the maximum  speed at which the nozzle can move during the execution of each movement command. The feed rate, often, can be set once for a sequence of G-code instructions that describe a small trajectory.

\begin{figure}[h]
    \centering
    \includegraphics[width=1\linewidth]{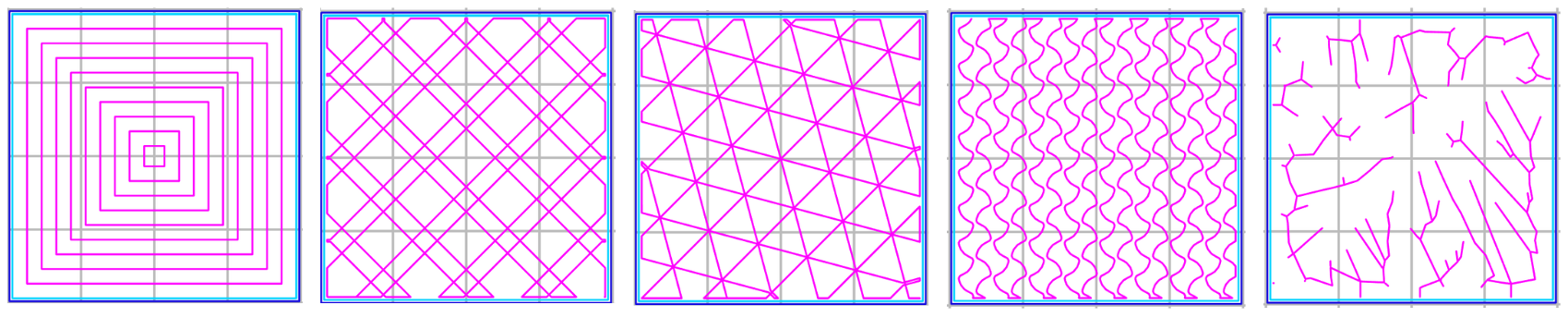}
    \caption{\textbf{Infill Patterns:} From left to right, the infill patterns are concentric, octet, triangle, gyroid, and lightning.}
    \hrulefill
    \label{fig:infill}
\end{figure}

Prior to slicing, there are several factors that are adjustable, all of which result in a different G-code and a different trajectory followed by the printer nozzle, even when the object printed is visually identical. There are three factors that are commonly adjusted/set by the designer. First, the 
 \textit{infill design}, which is the pattern of printing the object's interior is chosen based on the desired  level of strength and flexibility the object should have. Some infill designs are shown in \autoref{fig:infill}. Second, the 
\textit{infill density} describes how densely packed the infill design should be and is chosen based on how strong and/or heavy the final object needs to be. The last factor, \textit{translation},  refers to the position on the printing plate. While the default setting is the center of the printing plate, the user can customize it to print anywhere on the printing plate.

\new{\heading{Oriented Bounding Box and Maximum-Overlap.} One of the goals of the reverse-engineering solution is to recover the trajectory of the nozzle from the video of the 3D printing process. Since the recovered 3D trajectory may not have the same centroid as the actual trajectory of the nozzle (ground truth), our checker (as we will see later in Section \ref{subsec:curve_checker}) will effectively compare the recovered trajectory with the ground truth layer-by-layer, i.e., compare a sequence of 2D trajectories. In order to do so, for each layer, we need to enclose the points that describe the recovered 2D trajectory within a closed polygon, ``shift'' it so that it has the same centroid as the trajectory in the corresponding layer of the  ground-truth, and there is  ``maximal overlap'' between the two trajectories. Then, we can use a standard distance metric such as DTW~\cite{Mueller_DTW} to compute the similarity between the two trajectories. 

In computer vision, Oriented Bounding Box~\cite{9996428} is a geometric algorithm that can contain a sequence of points within a bounding rectangle or square. The advantage of this algorithm is that it allows the orientation of a sequence of points in 2D space to be changed, without affecting the relative position of the points with respect to each other. In our curve checker, we use a variation of this algorithm as explained in \ref{subsec:curve_checker}. For aligning the two objects, we use the well-known problem formulation \textit{Maximum Overlap}~\cite{maximum_overlap} in computational geometry, which is used for determining the optimal translation of one shape to maximize its overlap with another.}

%% file: Sources/Sections/methodology.tex
\section{Methodology}
\label{sec:methodology}

In this section, we first describe our G-code reverse-engineering solution (Section \ref{subsec:methodology_reverse_engineering_model}) that an adversary can use to reverse-engineer the 3D printable G-code IP from the video recording of the 3D print process of a target object. Next, in Section~\ref{subsec:curve_checker}, we present our 3D print-specific G-code equivalence checker that can detect layer-by-layer dissimilarity between two G-codes and can be used to evaluate the accuracy and feasibility of any G-code IP-theft attack.

\subsection{G-code Reverse Engineering}
\label{subsec:methodology_reverse_engineering_model}
The objective of the reverse-engineering model is to map the video from a print process to the corresponding G-code that describes the nozzle's movements across the build plate using G0/G1 commands. It also calculates the rate of filament extrusion per instruction and feed rate (\new{using } E and F \new{commands}). Recall that a 3D print process typically comprises several small movements of the nozzle, where each movement is due to a single G-code instruction. 

A single G-code instruction is composed of the command (usually G0 or G1), the next position of the nozzle on the print plate in the form of (X,Y,Z) coordinates, the extrusion rate (E), and the feed rate (F). We use a data-driven approach to predict the type of movement (G0 or G1) and the destination position (X,Y,Z) of the nozzle. We can then use deterministic approaches to calculate the remaining parameters (extrusion rate and feed rate).

\begin{figure}
    \centering
    \includegraphics[width=\linewidth]{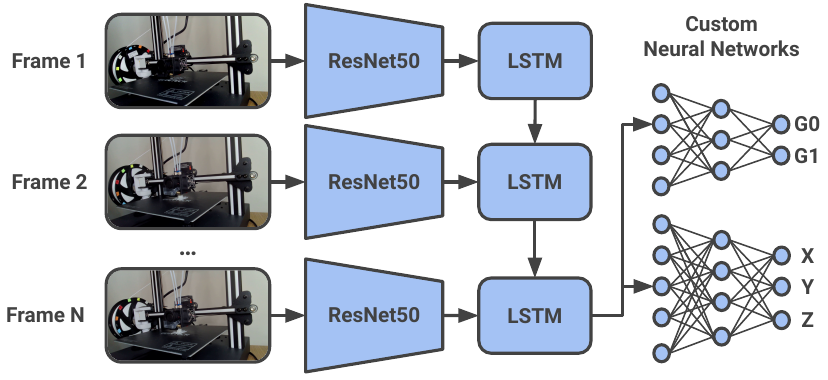}
    \caption{\textbf{G-code Reverse Engineering Model.} The model uses a ResNet-50 to extract features from individual frames of the input video, which is a chunk of the video recording of the entire 3D print. The features act as input to an LSTM that generates an embedding representing the whole video-chunk. This embedding is used as input to two custom neural networks that predict the movement action-command being executed (G0 or G1) and the destination coordinates of the nozzle.}
    \hrulefill
    \label{fig:re-lstm-model}
\end{figure}

\heading{ML-model architecture.} 
To obtain the position of the nozzle, our model uses $N=30$ frames as input, sampled from the video of a single movement of the nozzle (corresponding to a single G-code). For each frame ($224 \times 224$ image), we use ResNet-50 to extract its features, which serves as input to a Long Short-Term Memory (LSTM) network. The LSTM network captures the context, time, and 3D printer physics information of these frames, and generates a single embedding that represents the video. As we can observe in~\autoref{fig:re-lstm-model}, with ResNet-50 and LSTM we build two models (independent from each other) that can predict (i) the command corresponding to that movement, i.e., a binary classification model that predicts if it is a G0 or G1 command; and (ii) the corresponding coordinates of that G-code instruction. 

\new{Note that ResNet-50 was used by Liang et al. \cite{liang2022hiding} to map individual frames in the video to points in the 3D plane of the printer. Combining ResNet-50 with LSTMs allowed us to create a vector embedding of the 30-frame video chunk that loosely captures both the nozzle movement as well as the extrusion timing information. Also, this model combination allows us to efficiently process the video data, while being lighter than vision language models (VLMs)~\cite{vivit-vision-transformers} with respect to computational cost and training data requirements. We chose a frame rate (30 frames) that captures sufficient information without overwhelming the model with excessive data, leading to shorter training and processing times.}

In the \textit{command-prediction} model, the challenge is the class imbalance. For a single cube object, about 66\% of the commands are G1s, while only 33\% are G0s. When considering multiple G-codes from our dataset, the imbalance is greater: only 12\% of the commands are G0s. Due to that, we used a downsampling strategy to make the model unbiased towards the majority class~\cite{10.1145/3617897} (i.e., the model may learn that it is easier to classify everything as a G1 to increase the accuracy instead of learning the patterns of both classes), training it on the same number of G1 and G0 commands. Since it is a classification task, the custom neural network built on top of the LSTM for this task is trained using cross-entropy loss~\cite{10.5555/3327546.3327555}.

In the \textit{coordinates-prediction} model, the custom neural network on top of the LSTM is trained using the Mean Squared Error (MSE) loss function, given that the objective is to approximate the predictions to the ground truth as a regression task. In addition, in the output of the network, we use a sigmoid activation function multiplied by 250, which is the maximum value for all the axes of our 3D printer, thus limiting the range of each axis predicted by the model. In both models, we use Adam Optimizer~\cite{kingma2017adam} with a learning rate of $10^{-5}$, and Cosine Annealing with Warm Restarts~\cite{loshchilov2017sgdr} as scheduler ($\text{warmup epochs} = 40$), as most prior works do for video classification~\cite{arnab2021vivit, tong2022videomae}, and $32$ as batch size.

\begin{figure}
    \centering
    \includegraphics[width=\linewidth]{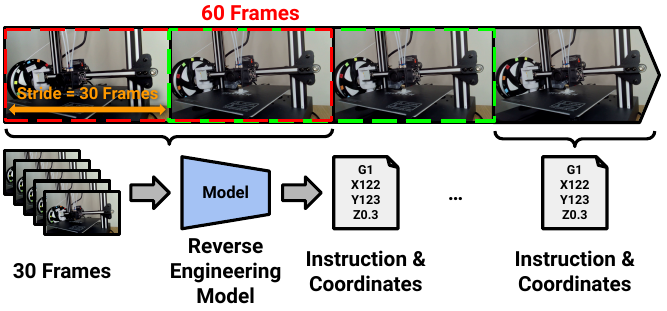}
    \caption{\textbf{Model inference using a sliding-window strategy.} Since the attacker has no knowledge of when an instruction is starting and ending, we used a sliding window strategy to obtain the G-code from a full video of a 3D printing process.}
    \hrulefill
    \label{fig:inference}
\end{figure}

\heading{Sliding Window Inference Strategy.} 
After training both models, the main challenge for an attacker is to make the inference and get the G-codes for a full video of a printing process because he has no knowledge of where instructions start and end, i.e., the video is not segmented by G-code instruction as in the training dataset. To solve this issue in a realistic way, we assumed this scenario and implemented a \textit{sliding-window} strategy to make the inference of a full video, as we present in~\autoref{fig:inference}. Given a window (which we refer to as $\mathrm{batch}$) containing 60 frames (in red), we sample 30 frames from that, present them to the models, and get their corresponding commands and coordinates. After that, we slide the window by 30 frames ($\mathrm{stride}$ of 30 frames in orange) and make the inference of the new window (in green). We repeat this process until we reach the end of the video (that has $n_{\text{frames}}$), resulting in a set of commands and coordinates that represent the full G-code of a single object.

\heading{Normalizing the Z-Axis.} Another challenge when reconstructing the G-code for the recovered object is to define the correct value for the Z-axis. Z-axis values are crucial in segregating the instructions into layers. If our model recovers each instruction accurately, but fails to recover, or recovers the wrong corresponding Z-axis value, we would have failed to carry out a successful IP theft.  Given that the model performs a regression task, the value is typically a floating-point number, but the printer uses a discrete interval to go through the layers (in our case, steps of 0.3 mm). In addition, we need to consider that the model output may be noisy and state that an instruction from a given layer is in another layer. To estimate when there is a proper change in the Z axis, we used the Pruned Exact Linear Time (PELT) change point detector algorithm~\cite{Killick_2012}, which identifies abrupt shifts or transitions in a time series or sequence of data, marking these locations based on underlying properties. We consider that the attacker can analyze the predicted Z coordinates and select the best parameters (penalty cost, and minimum length of segments) for a given 3D object. Finally, we initialize Z as being 0.3, and for every change point detected by PELT, we increase it by 0.3 (layer height), discretizing the Z axis for the 3D printer so it can properly print the object and effectively recreate the same structure of the ground-truth G-code generated by any slicer (e.g., CuraEngine).

\heading{Obtaining the Extrusion and Feed Rate.} 
\label{subsubsec:obtaining_extrusion_feed}
Once we have the command (G0 or G1) and coordinates (X, Y, and Z) from a data-driven approach (reverse-engineering model), we can obtain the extrusion rate and feed rate in a deterministic fashion. The extrusion rate describes the position of the filament in terms of input to the extruder feeder. It is calculated using the equation described in Section \ref{Section: Background}. 
The feed rate sets the maximum speed for the nozzle travel along X,Y, and Z axes and indirectly sets the rate of extrusion. 
That speed must align with the rate of extrusion, such that the nozzle does not travel faster than the filament extruded. The optimum speed setting is specific to both the 3D printer and the type of command used. In our experimental setup and dataset, most G0 commands have a feed rate of $7740$, while G1 commands have a feed rate of $3600$. Thus, based on the label given by the command prediction model, we can estimate the value of the feed rate $F$.

In summary, the reverse-engineering model recovers the printhead coordinates and the corresponding commands (G0 or G1) and thereby reconstructs a set of points that describe object topology, including its infill pattern and density. We then reverse-engineer the extrusion rate to calculate how much additional filament the printer needs to disperse between two adjacent points (in the case of a G1 command) and the feed rate to calculate the movement speed, resulting in the full G-code.

\subsection{G-code Equivalence Checking}
\label{subsec:curve_checker}
\input{Sources/Sections/Equivalence-Checker_Curve}

%% file: Sources/Sections/Equivalence-Checker_Curve.tex
We developed a G-code equivalence checker, referred to as the \textit{curve checker}, to assess the performance of our G-code reverse-engineering model. The curve checker compares the model's output G-code with the original or target G-code, identifying layer-by-layer dissimilarities between the objects represented by both codes. It also generates a comprehensive dissimilarity score by aggregating these layer-wise differences and applying a penalty if the target object's height deviates from that of the source object.

From the perspective of the attacker, there is no way to know the placement of the camera with respect to the coordinate frame used by the honest entity. As a result, the G-code recovered by our model can be rotated or translated with reference to the honest entity's coordinate frame. To that effect, our curve checker has to be invariant to rotation or translation. Throughout our comparisons, we will refer to the G-code with respect to which we are comparing as the ground truth (GT).

\heading{ Skeletal Structure.} The G-code reverse-engineering model recovers the trajectory followed by the printer nozzle during the printing process, in the form of a G-code. While G-codes mostly follow a set structure, there are several structural variations, which result in the same action by the 3D Printer. For example, the G-code instruction lines ``G0 X96.42 Y122.08'' and ``G1 X96.42 Y122.08'' perform the same action, which is the printer nozzle moving to the coordinate (96.42,122.08) on the build plate. Thus, our first challenge was to extract the trajectory followed by the nozzle during the printing from the syntactical structure of a G-code. We call this \textit{reducing the G-code to its skeletal structure}. 

A first step to solving this challenge was to understand the correlation between a G-code instruction and the corresponding nozzle action. For example, both the G0-command and the G1-command-without-extrusion have the same effect, which is nozzle movement without extrusion. However, both have different purposes. During a print, if there is no direct path for the nozzle to travel to the next coordinate, it retracts and moves around the semi-finished object to reach that point, which is encoded using G0. On the other hand, the G1-command-without-extrusion is used to encode the printer nozzle moving into position to begin extruding. Thus, instructions containing G0 commands can be excluded while excluding G1-without-extrusion commands can result in us having an incomplete printing trajectory. 

Using such correlations, we created a pattern for print trajectory commands. Recall the structure of a G-code from Section \ref{Section: Background}. Once we identify the print commands, we have two tasks: extract the X and Y coordinates as a tuple to form the 2D trajectory per layer and track the changes in the Z value to identify a change in layer. We achieve this using a nested list, also known as a list-of-lists, data structure~\cite{listsss}. When the Z-axis value changes, we begin a new list and keep appending (X,Y) values as a tuple, till the Z value changes again. Thus, the skeletal structure of a G-code is a nested list of tuples. The entire 3D object can be described by the outer list. A specific layer of the object can be described by the corresponding inner list.

\heading{Oriented Bounding Polygon.} Once we have the skeletal structure of each layer of the G-code, our next challenge is to preserve its vector nature, while we perform rotational and translational operations on it. Oriented Bounding Box (OBB) \autoref{Section: Background} is typically used for such tasks. \new{However,} OBB has a major drawback, especially in the 3D printing setup. \new{In OBB, two  unique trajectories that are described by a different sequence of points can be contained within the same bounding box.}

\begin{figure}[t]
\begin{subfigure}{0.22\textwidth}
\includegraphics[width=\textwidth]{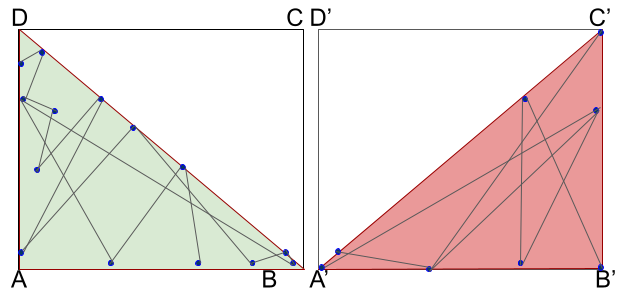} 
\caption{
}
\label{f1}
\end{subfigure}
\begin{subfigure}{0.12\textwidth}
\includegraphics[width=\textwidth]{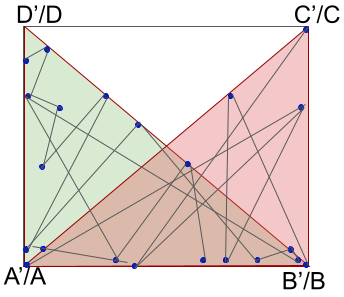}
\caption{
}
\label{f2}
\end{subfigure}
\begin{subfigure}{0.12\textwidth}
\includegraphics[width=\textwidth]{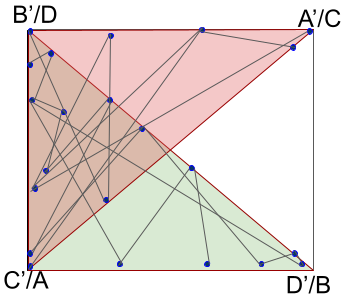}
\caption{
}
\label{f3}
\end{subfigure}
\caption{Visual representation of two unique trajectories contained within a bounding box of identical dimensions, and how the trajectories overlap when their bounding boxes are perfectly aligned. }
\hrulefill
\label{F1}
\end{figure}

Observe \autoref{f1}. ABCD and A'B'C'D' are two identical bounding boxes for two unique trajectories that we are attempting to align. According to the core principle of OBB, we would be aligning the boxes ABCD and A'B'C'D'. \autoref{f2} and \autoref{f3} show the only two positions where the bounding boxes are perfectly aligned. While there is a significant amount of alignment, there also exists a significant portion that is not aligned. \new{This limitation is primarily due to the fact that OBB only uses a rectangle or square to form the bounding box. To overcome this limitation, we created an algorithm, which is a specialized version of OBB, called Oriented Bounding Polygon~(OBP).}

\begin{figure}[t]
    \centering
    \begin{subfigure}{0.20\textwidth}
        \centering
        \includegraphics[width=\textwidth]{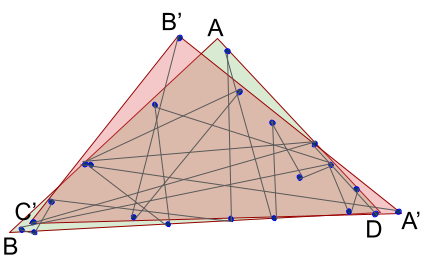}
        \caption{Perfect alignment of two Bounding Polygons .}
        \label{obp}
    \end{subfigure}
    \hfill
    \begin{subfigure}{0.20\textwidth}
        \centering
        \includegraphics[width=\textwidth]{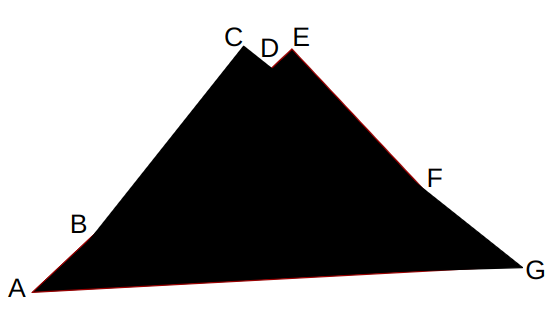}
        \caption{Fused polygon derived from  the Bounding Polygons}
        \label{fused}
    \end{subfigure}
    \caption{Bounding Polygons and Fused Polygons. \new{Observe that the alignment between the two trajectories is much more fine-grained compared to when we use a bounding-box to enclose the trajectories as shown in  Figure \ref{F1}}.}
    \hrulefill
    \label{merged_fig}
\end{figure}

In OBP, we use a \textit{convex hull}\new{~\cite{convexhull}} to create a tightly bound polygon to enclose the skeletal structure per layer. This overcomes the limitation of OBB, as arbitrary trajectories can now be enclosed tightly within an arbitrary polygon. For example, in  \autoref{f1}, we consider the triangles ABD and A'B'C' to be our bounding polygons instead of the bounding boxes. Now, in order to align these two polygons, we need to translate and rotate them in some way so that the two bounding polygons are maximally or perfectly aligned, for example as shown in \autoref{obp}. \new{Note that the algorithms~\cite{convexhull} for finding the convex hull typically use approximations that can potentially affect the final similarity scores between the reverse-engineered trajectory (RT) and the ground-truth (GT) that our curve checker reports. However, we did not see any anomalous results, i.e., curve checker reporting high similarity scores on dissimilar objects, and vice-versa, in our evaluation (Section ~\ref{subsec:reverse_engineering_model_eval}) that included commonly 3D printed objects of varying complexity. We verified this by plotting both RT and GT of all objects; see Figures \ref{keyy}, \ref{gear}.}


\heading{Perfect Alignment.} In our OBP algorithm, we can achieve perfect alignment through translation and rotation while freezing the enclosed points that define the skeletal structure of the trajectory of each layer. In the first step of our alignment function, we compute the centroid of both polygons. If the centroids do not coincide, we shift one polygon along X and Y axes till both centroids coincide. Next, we keep one polygon fixed, while rotating the other using the classical \new{\textit{Maximum Overlap}~\cite{maximum_overlap} problem formulation}. In our algorithm, we solve the maximum overlap problem in the following way. We first create a single, fused polygon by  treating the two polygons with a common centroid as one polygon. For example, if we were to treat the intersected polygons in \autoref{obp} as a single, fused polygon, it would appear like polygon ABCDEFG in \autoref{fused}. Then, we rotate one polygon by one degree to create a new fused polygon and calculate and record the area of the resulting fused polygon.  We repeat this process for different angles of rotation---~ one to 360. For each fused polygon, we compare the area of the ground truth with the area of the fused polygon.  We select the degree of rotation for which the area of the fused polygon is closest to the area of the ground truth to get the perfect alignment between the ground truth and the \new{recovered trajectory}. 

\heading{Curve Dissimilarity.} Once we have our perfectly aligned polygons, we focus on the final component of our algorithm, which is to calculate the dissimilarity between two curves. We have chosen our metric as Dynamic Time Warping (DTW)~\cite{bringmann2023dynamicdynamictimewarping}. DTW, as a distance metric, takes into account the overall shape of the curve, rather than individual data points. Additionally, it is invariant to the number of points that make up a trajectory. In other words, in the case of two trajectories, if one contains ten discrete points and the other contains a hundred discrete points, the dissimilarity would be no different from if they both contained ten discrete points.

In our algorithm, we use a variant of DTW called Subsequence-Aligned DTW ~\cite{Mueller_DTW}. This variant of DTW is capable of identifying a common subsequence between two trajectories and aligning them before computing the dissimilarity. This adds an element of additional alignment, which is effective when dealing with objects having multiple axes of symmetry (e.g., a cone). Finally, we normalize the dissimilarity such that the final score is expressed in the form of a percentage value between zero and 100 instead of absolute values, for ease of interpretation.

\heading{G-code manipulator.} While our curve checker is built to be rotationally and translationally invariant, we needed a dataset comprised of rotated and translated versions of an object to quantitatively evaluate this property. Note that the slicer software that generates G-code from a 3D model has in-built capabilities for rotating and translating the 3D model and generating its corresponding G-codes. However, slicers have two limitations. First, the slicer \new{(Ultimaker Cura~\cite{ultimakercura}) that we used for our evaluation} can produce G-codes for objects that are rotated in multiples of 15 degrees, making it impossible for us to be able to utilize all 360 degrees. Note that in our threat model, we do not make any \new{assumptions} about the camera angle. Second, \new{a} more crucial limitation is that, when the slicer is used to translate or rotate an object, the infill pattern does not translate or rotate in synchronicity. The infill pattern is shifted in its own way, resulting in a different object altogether. As a result, a perfectly aligned object will be slightly dissimilar. 

\begin{figure}[h]
\begin{subfigure}{0.24\textwidth}
\includegraphics[width=\textwidth]{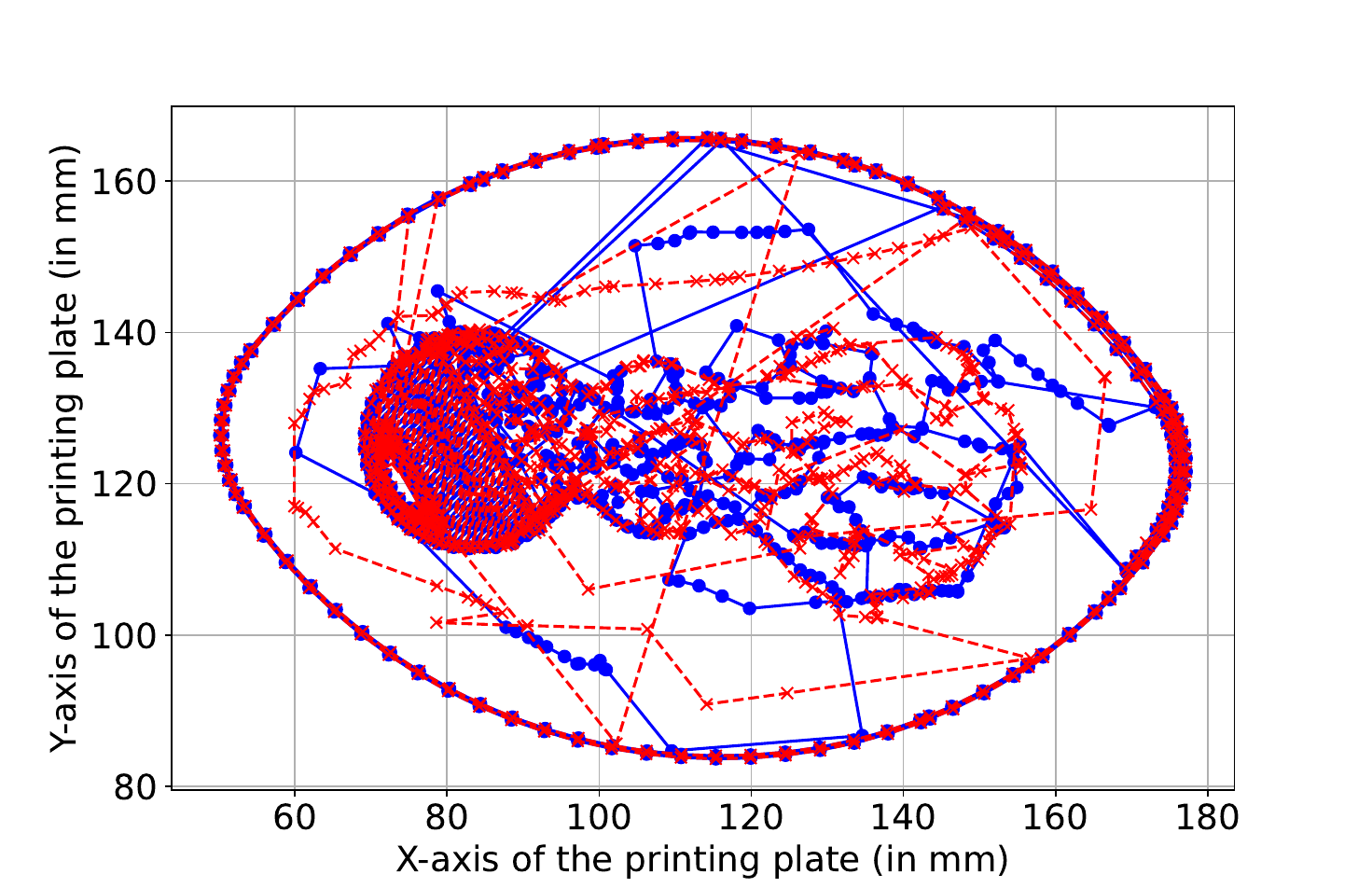} 
\caption{Rotated by the slicer \\ Dissimilarity = 3.16\% }
\label{fig:subim1}
\end{subfigure}
~
\begin{subfigure}{0.24\textwidth}
\includegraphics[width=\textwidth]{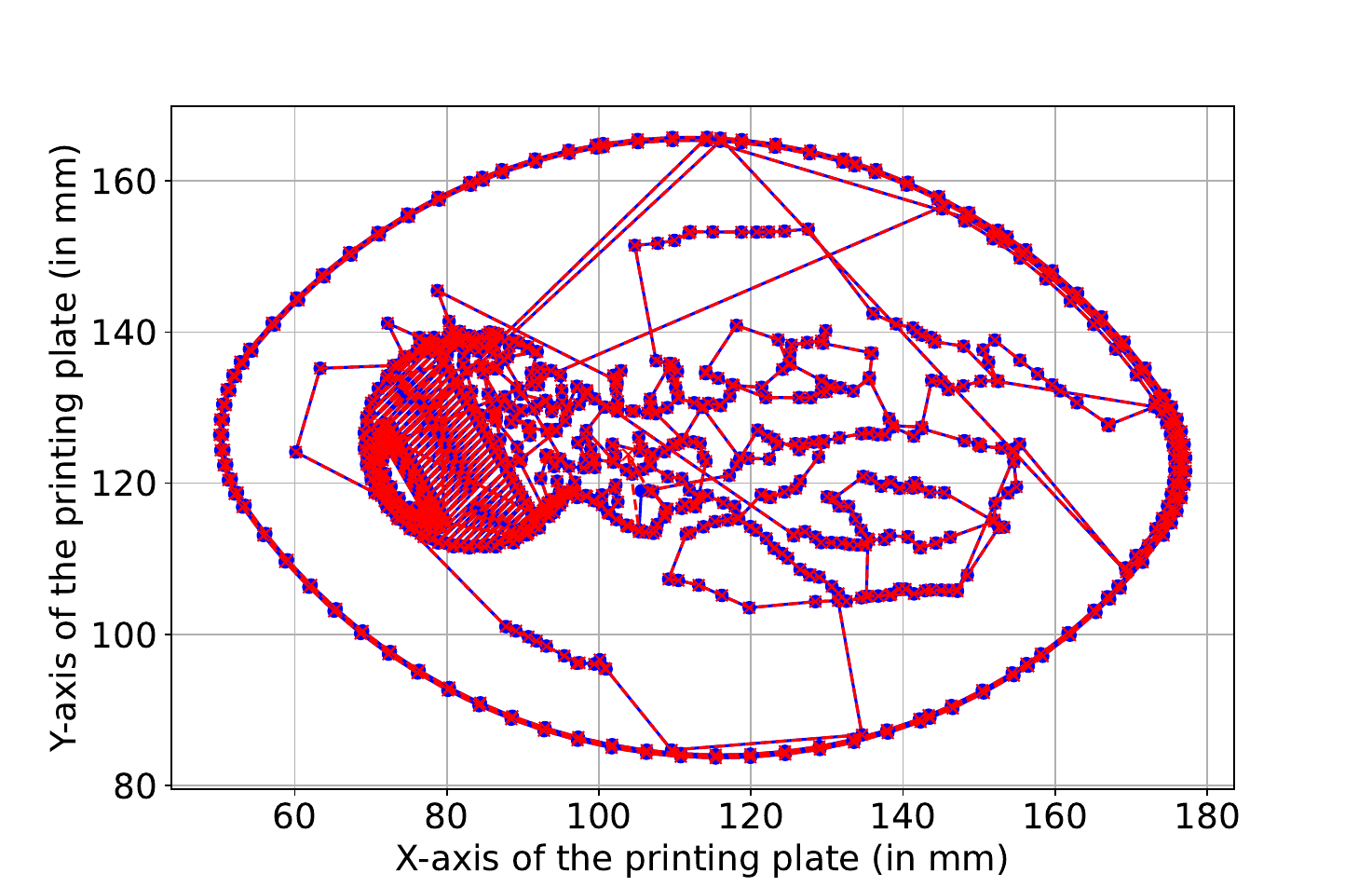}
\caption{Rotated by G-code Manipulator \\ Dissimilarity = 2.74e-16\%}
\label{figgy}
\end{subfigure}

\caption{Layer 79 of S4 (given in \autoref{obj}).}
\hrulefill
\label{fig6}
\end{figure}

In our experiment, we rotated an object by 30 degrees using the slicer. After perfectly aligning the original and the rotated object, we observed visible differences in the infill pattern, as shown in \autoref{fig:subim1}. Further, the curve checker reported a dissimilarity of 3.16\% between the two G-codes. Note that this dissimilarity can be higher for larger degrees of rotation. 
Thus, if we use the slicer to generate a dataset to evaluate our checker's invariance properties, we would be unable to distinguish our checker's limitation from the slicer's limitations, based on the values. 

To ensure that our curve checker is actually robust to rotation and translation, we built a tool called \textit{G-code manipulator} which takes a G-code and user-defined rotational (in degrees) and translational parameters (in mm), and produces rotated/translated versions of the objects without changing the infill pattern. It uses geometrical properties of shifting a curve along the X and Y axes for translation and applies a rotational matrix around the centroid for rotation. \autoref{figgy} shows the exact curve in \autoref{fig:subim1}, when rotated 15 degrees using our manipulator. We can see that the curves align completely, and our curve checker reports a negligible difference \new{(2.74e-16\%)} between the rotated object and the ground truth.

%% file: Sources/Sections/evaluation.tex
\section{Evaluation} 
\label{Section: Evaluation}
\label{subsec:setup}

\begin{figure*}
    \centering
\includegraphics[width=\linewidth]{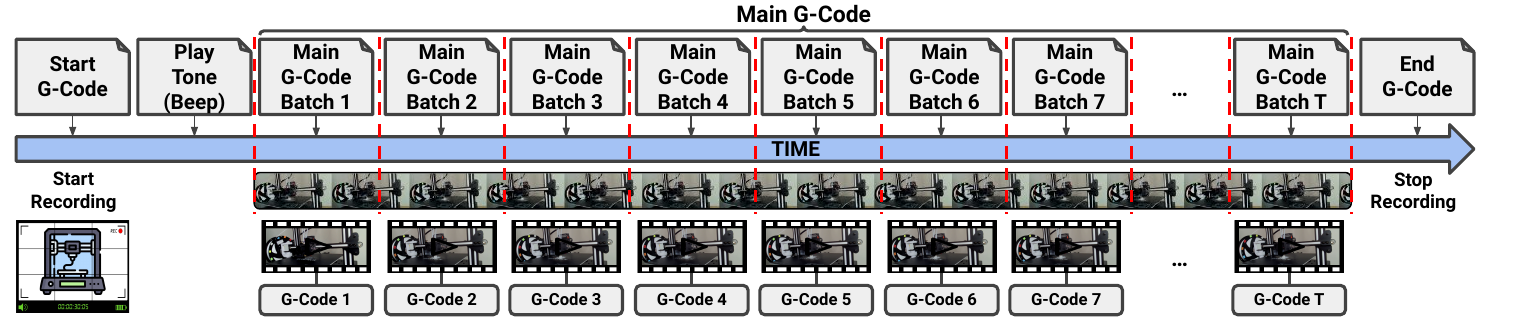}
    \caption{\textbf{The Data Collection Pipeline.} Our strategy consists of sending the start G-code, followed by a tone, which is used to sync the video with the instructions timestamps. Then, the recording starts and all the batches of G-codes are sent in sequence to the printer, mapping each batch to its corresponding video. After the figure is done printing, the recording is stopped and the ending code is sent to complete the print process.}
    \label{fig:data_collection}
    \hrulefill
\end{figure*}

The 3D printer used in our evaluation is Geeetech A20T~\cite{geeetech_a20t_manual}, with the firmware Marlin version 1.1.8~\cite{marlinfw}. It uses PolyLactic Acid (PLA) filament, has a $250$ mm $\times$ $250$ mm bed, and a $0.4$ mm nozzle diameter. The printing accuracy of this printer is $0.1 \approx 0.2$ mm, and the positioning precision for the X and Y axes is $0.011$ mm and $0.0025$ mm for the Z axis. According to the user manual~\cite{geeetech_a20t_manual}, Repetier-Host 2.3.2~\cite{repetier} is the default slicing software for this printer, including the best configuration files for the CuraEngine 15.01 slicer~\cite{ultimakercura}, with a print speed of 49 mm/s, outer perimeter speed of 43 mm/s, and infill speed of 78mm/s. We used concentric lines as the default infill pattern to 3D print objects because they produce the same infill pattern no matter where the object is in the bed. Finally, we used $0.3$ mm as both the first-layer height as well as the  intermediate-layer height.

\subsection{Data Collection}
\label{subsec:data_colection}
In this section, we describe the two datasets that we collected for the evaluation of our developed solutions for analyzing G-code IP-theft in additive manufacturing. Our G-code reverse-engineering  model shows how an adversary can reverse-engineer/steal the G-code IP of a manufacturer using just the video recording of the print process and our curve checker shows how close the stolen G-code can be to the manufacturer's G-code.

\heading{\new{G-code to video dataset\#1.}} 
\label{subsubsec:video_dataset}
To evaluate our reverse-engineering model, we created, to the best of our knowledge, the largest existing G-code-to-video dataset in literature, with 37,121 videos and their corresponding G-code instructions. In total, our dataset has more than 150GB of videos \new{(roughly 48 hours in total)}, consisting of 16 different objects whose 3D models we downloaded from repositories such as Thingiverse~\cite{thingiverse}, including variations of gears, keys, etc. Note that our choice of objects was limited by the fact that 3D printing takes hours to print even simple objects, especially in our \textit{batch-wise} print setup that we describe next.
\begin{table}[!ht]
\small 
\centering
\begin{tabular}{@{}l l r r@{}}
\toprule
\textbf{Set} & \textbf{Object (Acronym)} & \textbf{Instructions} & \textbf{Frames} \\ \midrule
\multirow{9}{*}{Train} 
 & Cube \twisha{(SOC)} & 1,870 & 116,330 \\
 & Hexagon \twisha{(HEX)} & 1,970 & 124,347 \\
 & Pyramid  \twisha{(PYR)}& 1,530 & 97,549 \\
 & Cone \twisha{(CON)} & 6,250 & 350,561 \\
 & Hexagonal Pyramid \twisha{(HPY)} & 1,250 & 78,683 \\
 & Hexagonal Leveling Foot (HLF) & 973 & 60,055 \\
 & Hexagonal Grid Alternative \twisha{(HGA)}& 2,011 & 125,352 \\
 & Gear 19 Tooth \twisha{(G19)} & 4,742 & 276,517 \\
 & Key Five \twisha{(KE5)} & 3,458 & 212,282 \\ \cmidrule{2-4}
 & \textbf{Total} & \textbf{25,027} & \textbf{952,877} \\ \midrule
\multirow{7}{*}{Test} 
 & Tetrahedron (TET) & 1,020 & 48,817 \\
 & Dodecahedron (DOD) & 820 & 43,476 \\
 & Hexagonal Pyramid Truncated (HPT) & 880 & 43,545 \\
 & Hexagonal Grid (HEG) & 1,510 & 75,326 \\
 & Gear 16 Tooth (G16) & 4,010 & 162,640 \\
 & Key Four (KE4) & 3,104 & 131,121 \\
 & Hollow Cube (HOC) & 750 & 39,300 \\ \cmidrule{2-4}
 & \textbf{Total} & \textbf{12,094} & \textbf{544,225} \\ \bottomrule
\end{tabular}
\caption{\textbf{Dataset\#1 Distribution (Number of Instructions and Frames) \new{for training our base model}}. In total, we have more than 37K instructions and almost 1.5 million frames. \new{We used a subset of dataset \#1 to show the robustness of our solution to different camera angles.}}
\label{tab:gec-objects}
\end{table}

One of the major challenges in creating this dataset was to synchronize the video chunks with the corresponding G-code instructions. The shallow queue implemented in the native Marlin firmware \cite{marlinfw} introduced an unknown, varying delay between timestamps recorded by the Repetier-Host software~\cite{repetier} and G-code command execution by the printer. Instead of directly sending the print commands to the firmware, we used  OctoPrint~\cite{octoprint} to send the G-code instructions to the printer using the OctoPrint REST API while simultaneously monitoring the prints using the Python OctoRest library~\cite{octorest}. This setup allows us to save the timestamps when a printing process starts and ends. 
Another challenge that we faced was the significant delays in sending G-code instructions one by one to the printer via Octoprint; this resulted in mangled 3D prints as the temperature of the nozzle and the bed changed due to the wait time.  
We overcame this challenge by splitting the full G-code into batches of $N$ instructions and uploading them to the printer as separate files. 

Each print job, defined by the batch of $N$ instructions, is wrapped by two special instruction sets at the beginning (to set the bed and nozzle temperature) and at the end (to cool down the nozzle and move it back to the base position) to set up the 3D printer for each batch print.  
We then begin printing the object in batches, saving the timestamps of each batch execution, thereby making sure that each video segment captures precisely $N$ G-code instructions. Finally, we terminate the recording and execute the ending G-code to finish the printing process. Refer to~\autoref{fig:data_collection} for a visual high-level overview. In this work, each batch has $N=1$ G-code instruction.

To synchronize the beginning of the print process with recording without accessing the printer firmware, we initialize each print with a manually inserted \textit{M300} command that plays a tone right after the start G-code finishes. Our solution uses the tone's signal to map its corresponding G-code to the correct video frame. 

\new{\heading{G-code to video dataset\#2.} In order to test the \textit{robustness} of our reverse-engineering solution, we collected data from a second G-code to video configuration where the camera was placed at an angle of $60^{\circ}$ clockwise relative to the camera position used in dataset\#1. We 3D printed  3 objects (PYR, HPY, and SOC) that resulted in an additional dataset of 5 hours of 3D-printed videos and their corresponding G-code instructions.}

\heading{Dataset for evaluating curve checker.}
\label{subsubsec:gcode_dataset} 
To evaluate our curve checker, we created a dataset of objects that are commonly printed in departments across the 16 critical infrastructure sectors ~\cite{critic}. The G-codes used in our evaluations have been taken from open source platforms (e.g., Thingiverse~\cite{thingiverse} and GrabCAD~\cite{grabcad}). However, the objects themselves, given in \autoref{obj}, are available in the market with numerous customized and proprietary variations. We classified the objects into simple (S1-S6) and complex (C1-C6) based on the number of layers, the number of points per layer, and the overall geometry of each object. 

\new{ The number of layers in the simple objects does not exceed 120. Further, the time taken by the curve checker to evaluate all datasets of a simple object took an average of one hour, with the upper limit not exceeding 90 minutes. In contrast, the number of layers in the complex objects ranged between 300 to 900, with each object taking a minimum of 3 hours to be evaluated. Also, all the simple objects in our dataset are slightly more complex versions of basic shapes. For example, S1 can be seen as an elongated, L-shaped hexagon, S2 can be seen as a cylinder with spiral cuts, etc. In comparison, none of the complex objects can be classified into any \textit{one} simple shape. They are composed of multiple different basic shapes. }

\begin{table}[t]
\centering
\resizebox{0.45\textwidth}{!}{
\begin{tabular}{@{}clcl@{}}
\toprule
\textbf{Acronym} & \textbf{Object} & \textbf{Acronym} & \textbf{Object} \\ \midrule
S1 & Hex Wrench & C1 & Turbine Impeller \\
S2 & Drill Bit Plate & C2 & Ball Valve \\
S3 & Laptop Stand & C3 & Prosthetic Finger Joint \\ 
S4 & Bottle Opener & C4 & Satellite Dish \\
S5 & Handguard & C5 & Toroidal Propeller\\
S6 & Lidar Mount & C6 & Archimedes Turbine\\ \bottomrule
\end{tabular}
}
\caption{\textbf{Real-world objects used to evaluate rotational and translational invariance of our curve checker.} Note that since the invariance evaluation can be done using the slicer which can generate G-codes very fast, we included real-world complex objects in this evaluation.}
\label{obj}
\end{table}
\begin{table}[t]
\centering
\resizebox{0.45\textwidth}{!}{
\begin{tabular}{@{}lp{\linewidth}@{}}
    \toprule
    Dataset &  Description\\ \midrule
    R & The original object was rotated by 180 degrees, with a 5-degree interval, resulting in 36 G-codes per infill design and 144 G-codes per object.  \\
    T & The original object was translated to 36 points on the printing plate, resulting in 36 G-codes per infill design and 144 G-codes per object.  \\
    RT & The original object was translated to 8 points on the printing plate. At each point of translation, the object was rotated by 180 degrees, with a 5-degree interval, resulting in 288 G-codes per infill design and 1152 G-codes per object. \\ \bottomrule
\end{tabular}
}
\caption{\textbf{Datasets}}
\hrulefill
\label{dtsts}
\end{table}

Once we selected the objects for evaluation, we sliced the 3D models using Ultimaker Cura~\cite{ultimakercura} to create the G-codes. We created four versions of each object, each with a unique infill design. The four infill designs we chose were Concentric, Gyroid, Lightning, and Triangle, as can be visualized in \autoref{fig:infill}. We then used our G-code manipulator to generate three datasets, as shown in \autoref{dtsts}. 

\begin{figure}[h]
\begin{subfigure}{0.235\textwidth}
\includegraphics[width=\textwidth]{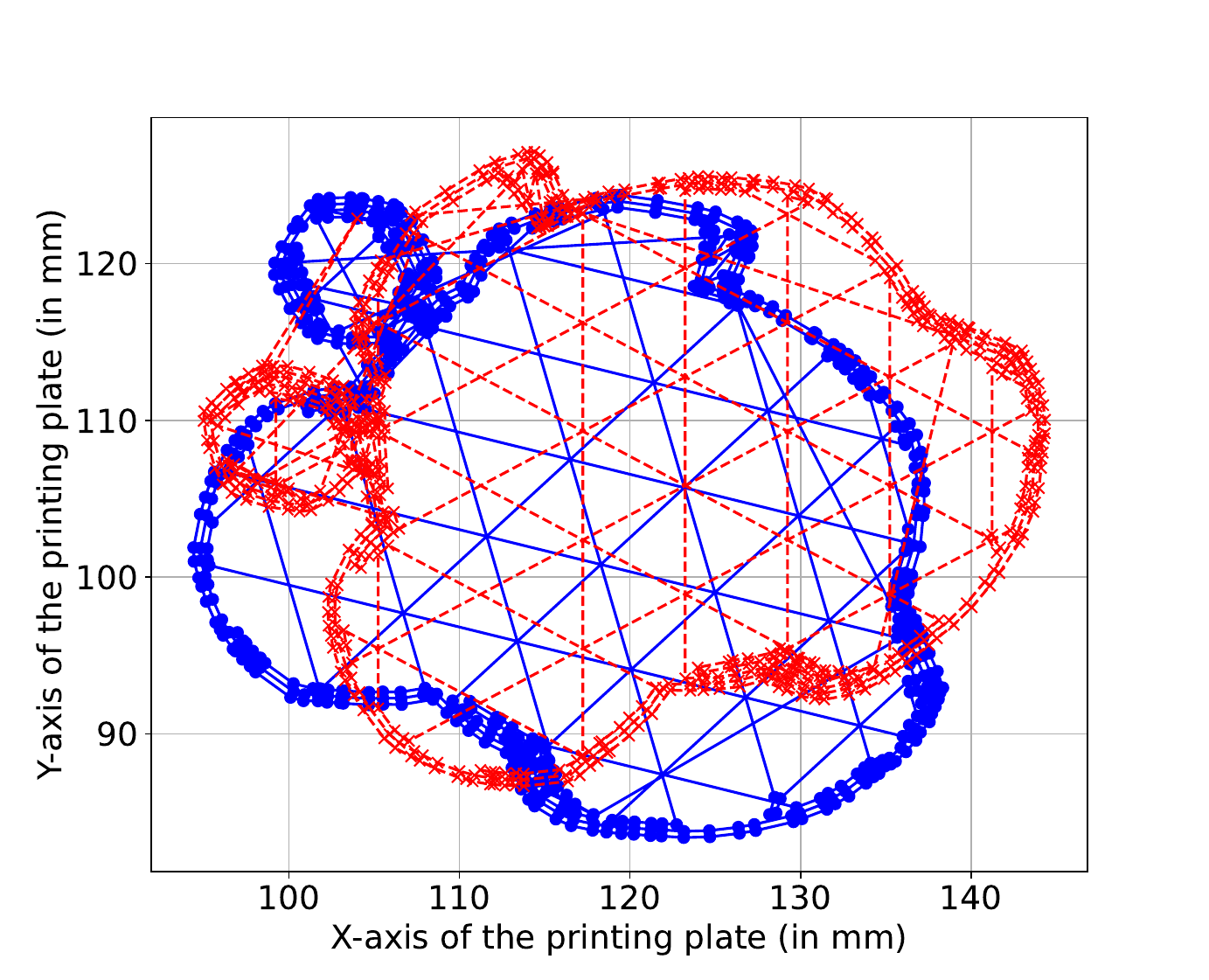} 
\caption{Rotated and Translated: Dissimilarity=7.03\%}
\label{fs1}
\end{subfigure}
\hfill
\begin{subfigure}{0.235\textwidth}
\includegraphics[width=\textwidth]{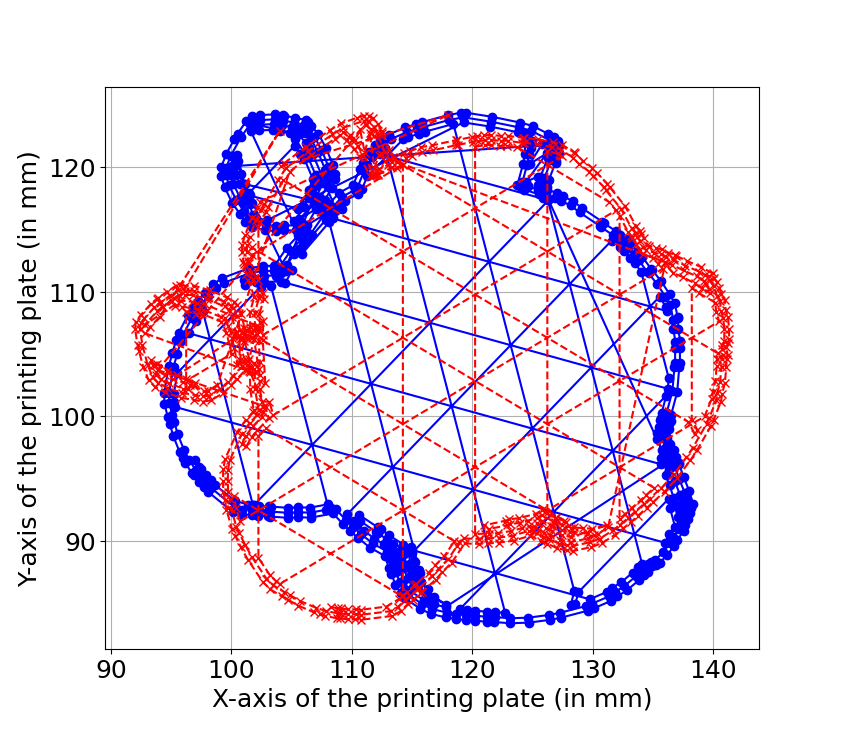}
\caption{Rotated, not Translated: Dissimilarity=6.49\%}
\label{fs2}
\end{subfigure}
\begin{subfigure}{0.235\textwidth}
\includegraphics[width=\textwidth]{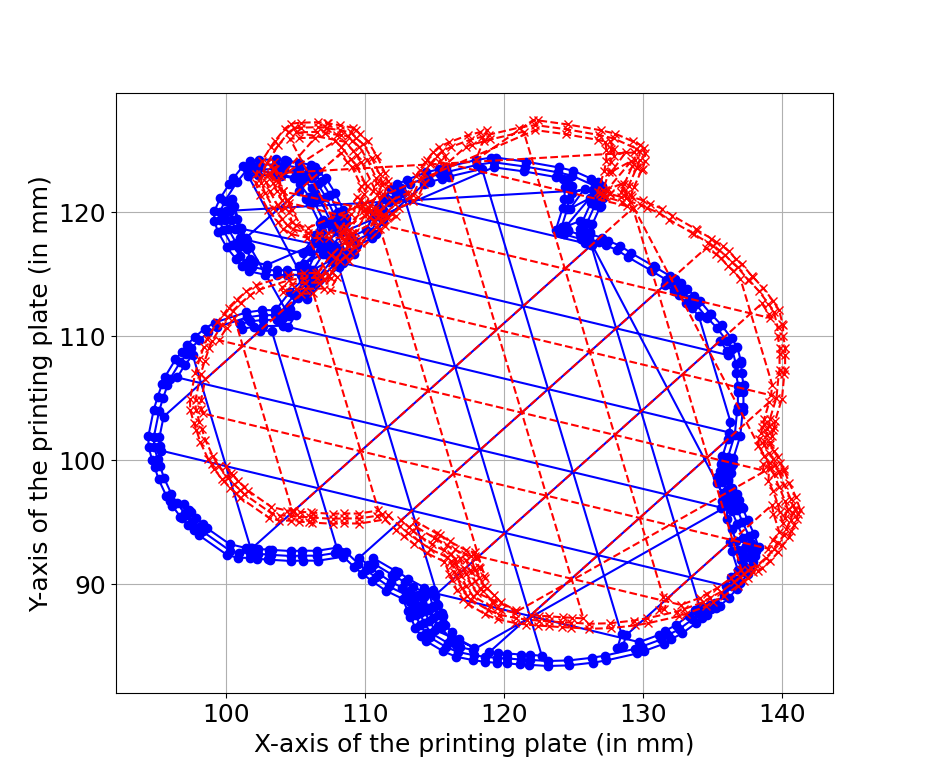} 
\caption{Translated, not Rotated: Dissimilarity=3.43\%}
\label{fs3}
\end{subfigure}
\hfill
\begin{subfigure}{0.235\textwidth}
\includegraphics[width=\textwidth]{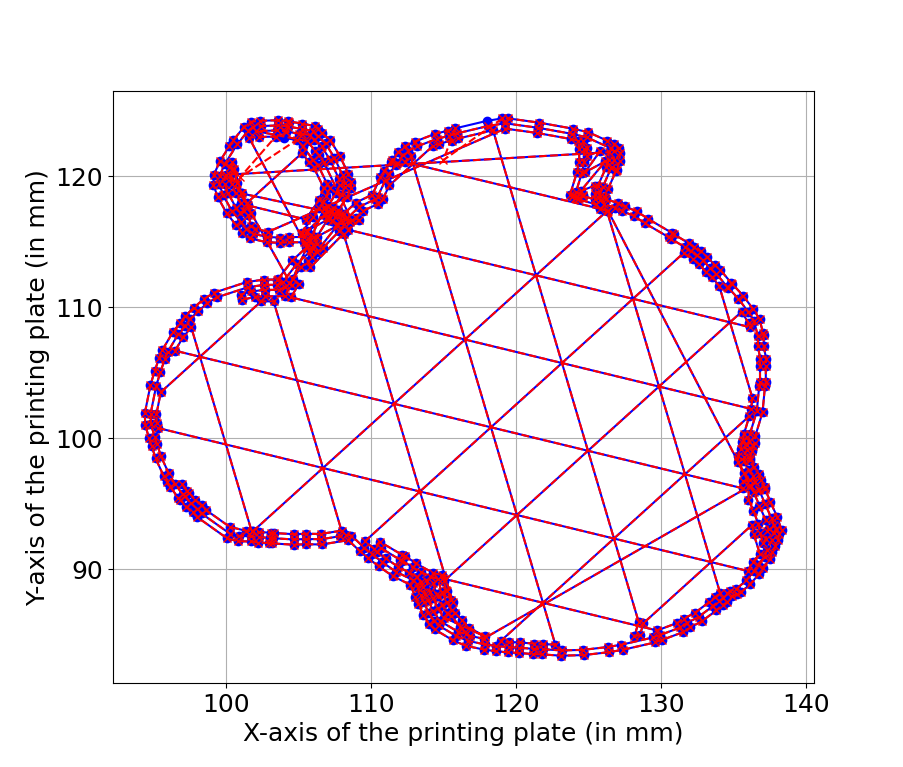}
\caption{Perfectly Aligned: Dissimilarity=0.01\%}
\label{fs4}
\end{subfigure}
\caption{All four images display Layer 840 of the Olympic Trophy.}
\label{oly}
\hrulefill
\end{figure}

To give an effective visual explanation of the datasets in \autoref{dtsts}, we chose an asymmetrical object, Layer 840 of an Olympic Trophy. In \autoref{oly}, blue curves represent the original trajectory, and red curves represent the recovered trajectory. \autoref{fs2} gives a visual representation of Dataset R. Both the original and the recovered trajectory have the same centroid, but the recovered trajectory has been rotated by 30 degrees counterclockwise around the centroid. We applied this logic in creating the R Dataset. Using the G-code manipulator, we created rotated versions of the original G-code at 5-degree intervals, through 180 degrees. 

The T Dataset is shown in \autoref{fs3}. In our visual example, the recovered trajectory has been translated by 4 mm along both x and y axes. To create our T Dataset, we first created a square grid with a spacing of 4 units. The center of the grid had coordinates (X,Y)=(0,0), and it was bound by the lines X=10, -10
and Y=10, -10. The original G-code had its centroid coincide with the center of the grid. Every G-code in the T Dataset had its centroid coincide with a lattice point on the square grid.

The TR Dataset applies the logic of both the T and R Datasets simultaneously. It is translated according to the logic of the T Dataset, and at every point of translation, it is rotated through 180 degrees, following the logic of the R Dataset. Each combination of translation and rotation results in a G-code, forming the TR Dataset. The visual of that can be observed in \autoref{fs1}, where the recovered trajectory has been rotated by 30 degrees and translated by a factor of 4 mm along both x and y axes.

Our datasets were created to demonstrate how the curve checker can perfectly align two curves, as shown in \autoref{fs4}, no matter their original orientation, and evaluate the true dissimilarity between two curves.

In summary, the evaluation for each object consisted of all three datasets, consisting of a total of 1440 G-codes per object. On an average, our curve checker took three hours to evaluate each simple object and eight hours to evaluate each complex object.

\subsection{G-code Reverse Engineering}
\label{subsec:reverse_engineering_model_eval}

As a proof of concept (PoC), we first trained our G-code reverse-engineering model on the cube presented in the training dataset from Section~\ref{subsubsec:video_dataset}. Based on the training loss, we observed that the model was converging and could successfully learn the cube's G-code. Then, we further trained the model to include all objects in \autoref{tab:gec-objects}, transferring the learning from the cube to all other 3D shapes. 

\begin{figure}[h]
    \centering
    \includegraphics[width=\linewidth]{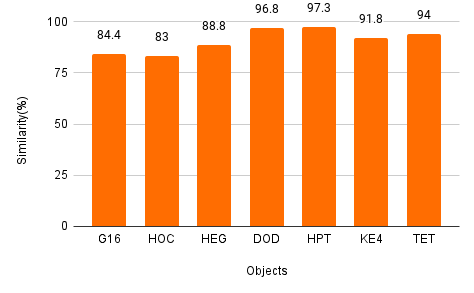}
    \caption{\textbf{Similarity between reverse-engineered G-code and the original G-code used for 3D printing objects (listed in~\autoref{tab:gec-objects}).}}
    \label{fig:gec-results-lstm-resnet}
    \hrulefill
\end{figure}

\heading{Accuracy.} First, we used our curve checker to evaluate the accuracy of the G-codes reverse-engineered by our model. Since the curve checker measures the similarity between two G-codes, we used the original G-codes that were used to print the objects as the second input to our checker. We present our results in \autoref{fig:gec-results-lstm-resnet}. Observe that we achieve a very high similarity percentage for our objects, with an average similarity of around 91\%. The highest accuracy of our model, as seen in \autoref{fig:gec-results-lstm-resnet}, is 97.3\%, with the lowest being 83\%. The values shown in the graph were averaged across all the layers. We conducted a deeper analysis on a per-layer basis. We noticed that the \new{largest} dissimilarity was in layer 1 \new{for most objects}. Layer 1, also known as the brim, is the layer that helps adhere the object being printed to the printing plate. It serves no further purpose and is peeled off once the object has cooled. The difference between objects with a higher similarity, like DOD, versus an object with lower similarity, like G16, is that G16 has fewer layers. DOD has 23 additional layers, compared to G16. Thus, the initial high dissimilarity in layer 1, which is irrelevant to the object itself, gets dissipated more in the case of DOD as compared to G16.

\heading{Counterfeiting via reverse-engineering.} For our functionality tests, we used a master padlock, whose key has a 4-number bitting, \new{and a 16-tooth gear, which is part of a larger gear system. 

For a key, the bitting }determines all relevant information regarding \new{its} geometry (each number corresponds to the depth of a cut on the key blade)~\cite{key_bitting}. We manually recorded the bitting of the key that opens the padlock and used keygen~\cite{keygen} to generate a 3D model with an equivalent bitting. \new{For a 16-tooth gear to be functionally-perfect-counterfeit,  it should be able to be the substitute for the original in a given gear system. For that, the teeth of the gear as well as its radius should match the original.} Next, we  sliced the 3D models \new{of the key and the 16-tooth gear} using our slicer to get the \new{G-codes and 3D printed both objects while simultaneously recording the print process.}

Since a real-world adversary knows in advance which object it is reverse-engineering, we further trained our model (which was already trained on a key KE5) on four random key designs. This is important to ensure that the attacker does \textit{not} have the same key in the dataset that opens the padlock, i.e., the model does not know what that key is, and also to ensure that our attack setup mimics a real-world adversary.

\begin{figure}
    \centering
    \includegraphics[width=\linewidth]{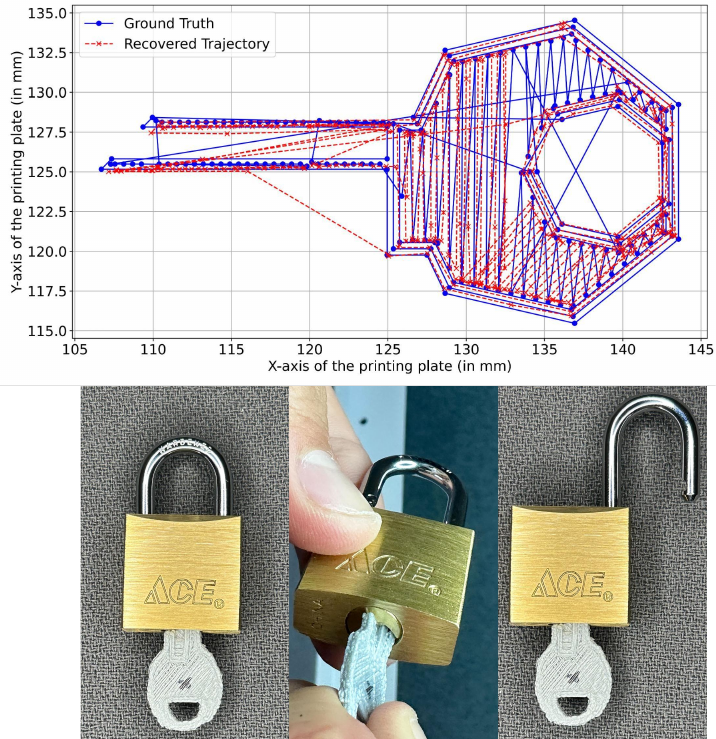}
    \caption{\new{\textbf{Top: Reverse-engineered key trajectory overlapped over the original key's trajectory in layer 3. Bottom: Functionality test of the counterfeit 3D-printed key.} The counterfeit key is 91.8\% similar to the original one and is able to open the padlock.}}
    \label{keyy}
    \hrulefill
\end{figure}

Given the video recording of the the key's printing process, our model outputs a G-code that is 91.8\% similar to the ground truth. We can observe in \autoref{keyy} that the main dissimilarity between the original and the reverse-engineered key lies at the top of the key, not the bitting, making it a \textit{functionally-perfect counterfeit}. We further prove our point by using our ``counterfeit'' key to open the padlock. As shown in \autoref{keyy}, we successfully open the padlock with the counterfeit key and demonstrate the \textit{functional correctness} of our model. We had previously reverse-engineered a key, KE4, with a 91.8\% accuracy (\autoref{fig:gec-results-lstm-resnet}) on a model trained with multiple different objects. While training the model on random keys did not increase overall accuracy, it allowed the model to identify the key features of the key. As a result, the key reverse-engineered was functionally accurate, compared to KE4, which had incorrect bitting.


\begin{figure}[!htbp]
    \centering
    \includegraphics[width=\linewidth]{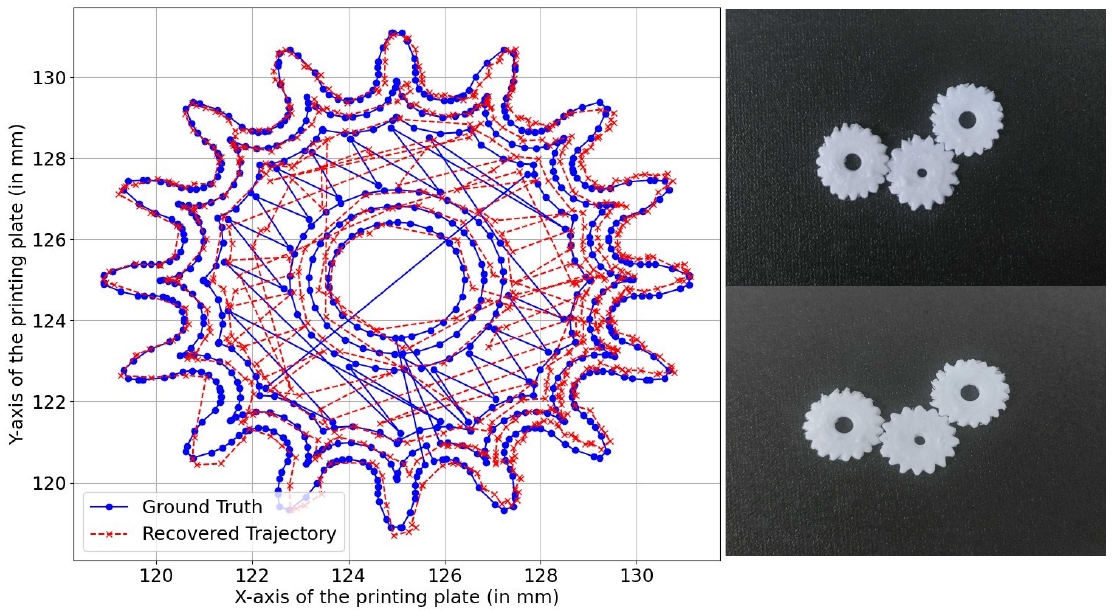}
    \caption{\new{\textbf{Left: Reverse-engineered gear trajectory overlapped over the original gear's trajectory in layer 3. Right: Functionality test of a counterfeit 3D-printed 16-tooth gear.} The counterfeit (middle) gear is 84.4\% similar to the original one and is able to fit into the three-gear system.}}
    \label{gear}
    \hrulefill
\end{figure}
\new{To demonstrate the functionality of the counterfeit gear, we printed a gear system, as shown in the \twisha{top right of} \autoref{gear}. We then substituted the middle gear with our reverse-engineered version in the gear system as shown in the \twisha{bottom right of} \autoref{gear}. Notice how in both figures, the radius and the teeth are identical, causing the reverse-engineered gear to perfectly fit into the system, demonstrating its functionality. Our curve checker reported an average similarity of $84.4\%$ with the maximum dissimilarity occurring due to a mismatch in the infill pattern which is a non-important factor for a functionality check in this case study analysis.}

\new{In both the counterfeit (key and gear) case studies, there were small differences between our reverse-engineered G-codes and the original G-code. However, both the counterfeit 3D-printed objects passed the basic functionality checks as the portions of the counterfeit objects that were dissimilar relative to the original objects did not affect the basic functionality. Therefore, we can say that our reverse-engineering solution is  able to reverse-engineer a G-code that is equivalent to the original from a functionality perspective. It might be the case that the counterfeit 3D-printed objects do not possess the same mechanical structural integrity properties as the original objects. In such cases, the owner of the G-code IP may define a set of properties that needs to be fulfilled in order to deem an object as counterfeit. However, such an evaluation is outside the scope of this work. }

\heading{Robustness to changes in camera angles.} As an attacker, we merely have an observational role regarding the video footage of the 3D printing process. As a result, we cannot exert any control over the placement of the camera. In order for our solution to be robust, we have to ensure that our attack cannot be deterred by simply changing the angle at which the honest entity is monitoring their print. \new{Given that the camera is part of the 3D printing setup, as stated in our threat model in Section~\ref{Section: Introduction}, for a given target, the viewing angle of a camera does not change typically.}

\begin{figure}
     \centering
     \begin{subfigure}[b]{0.22\textwidth}
         \centering
         \includegraphics[width=\textwidth]{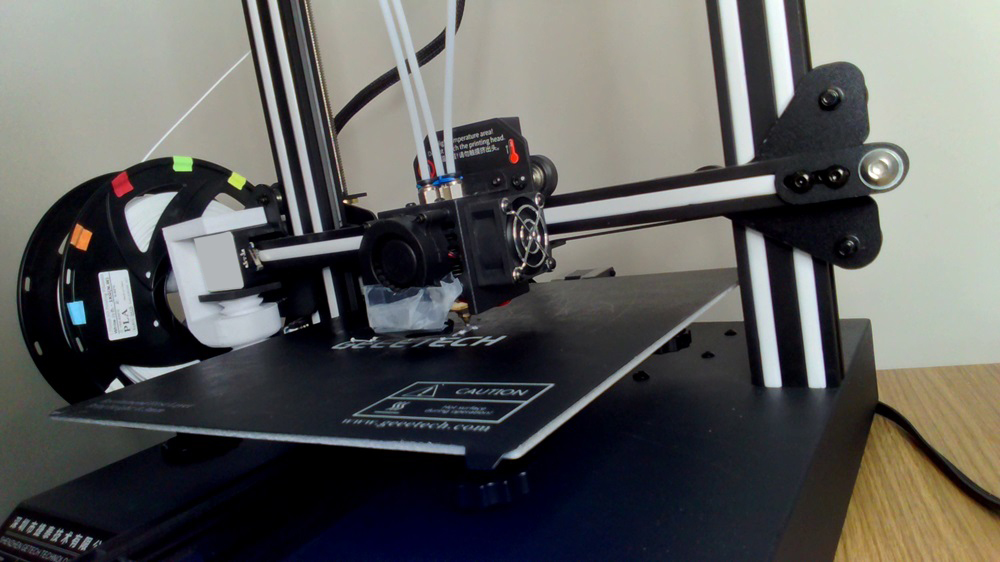}
         \caption{\textbf{Regular Camera Angle.}}
         \label{fig:regular_angle}
     \end{subfigure}
     ~
     \begin{subfigure}[b]{0.26\textwidth}
         \centering
         \includegraphics[width=\textwidth]{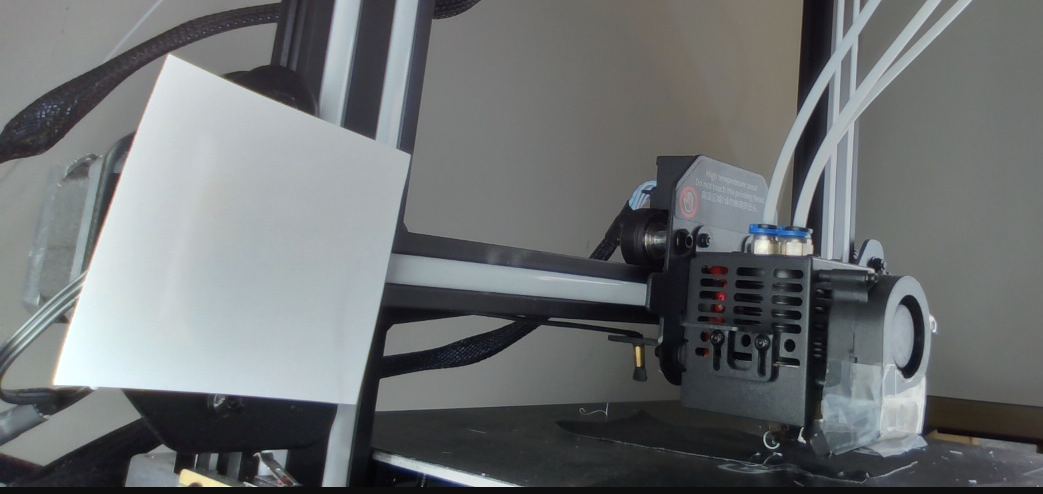}
         \caption{\textbf{\twisha{60-degree} Camera Angle.}}
         \label{fig:new_angle}
     \end{subfigure}
    \caption{\new{\textbf{Camera Angles for Creating G-Code to-Video Datasets.} Our main experiments used the camera angle shown in (a) to record the 3D printing process. To test the robustness of the model to camera angles, we changed the camera angle to $60^{\circ}$ relative to (a) as shown in (b).}}
    \label{fig:camera_angles}
    \hrulefill
\end{figure}

\new{

We evaluated the robustness of our reverse-engineering solution against changes in camera angles. As mentioned in Section~\ref{subsubsec:gcode_dataset}, we collected data from another G-code-to-video configuration with an alternative angle (dataset \#2) - see Figure~\ref{fig:camera_angles}. We trained our model using data from both angles and tested it on \textit{HPT} and \textit{HOC} objects in the G-code-to-video dataset \#2. We recovered HPT with an accuracy of 92.11\%, and HOC with a 98.33\% accuracy. The high accuracy shows the robustness of our reverse-engineering solution to change in camera angles. Note that in our threat model, the 3D printing process is remotely monitored via cameras before an attack occurs. Defenses such as injecting optical noise or adjusting lighting are not applicable as that would adversely affect legitimate monitoring by operators in a commercial setting.}


\new{\heading{Transferability to other 3D printers.} 
The model's core learning goal ---~to recover the printer nozzle's trajectory/extrusion---~ is 3D printer\footnote{\new{Our work is focused on filament-based single extruder 3D printers; other printers (e.g., metal-based) are out of scope.}} agnostic. Since the extrusion rate and feed rate can be derived from the trajectory and printer-specific information (e.g., extruder nozzle diameter), the parameter values for each G-code instruction for an adversary-owned 3D printer (e.g., Ultimaker~\cite{ultimakercura}) can be derived from the G-code instructions that were used to 3D print the target object on a different printer (e.g., Geeetech~\cite{geeetech_a20t_manual}). In short, the parameter values of G-code instructions are transferable from one 3D printer to another and we demonstrated this transferability in our functional evaluation of the counterfeit key where we recovered the G-code instructions using video-recording of the 3D-printing of keys in a Geeetech printer, and successfully 3D-printed its counterfeit version on  an Ultimaker printer.
}


\subsection{Comparison with prior work}
 There are multiple solutions that use side-channel attacks to reverse-engineer G-code, which we discussed in detail in Section \ref{sec:related-work}. However,  Liang et al.~\cite{liang2022hiding} model is the only one using an optical side-channel attack, and operating under a threat model similar to the one described in Section \ref{Section: Introduction} to reverse-engineer a G-code. They also use a distance-based metric, Mean Squared Error, to evaluate the accuracy of their model. As a result, Liang et al. model is the model we selected to compare our results with.

 \new{Before presenting our results on the comparison  between our work and Liang et al~\cite{liang2022hiding}, we want to reiterate that the threat models of both are different. While Liang et al. assumed that the optical side channel can be obtained from cameras that are either planted by an adversary or are part of the existing 3D-print monitoring/surveillance setup. We assumed the latter as it is more realistic compared to an adversary getting access to a commercial 3D-printing factory. Note that Liang et al.'s proposed defense of noise injection is not applicable to our threat model as that will also affect the quality of the monitoring/surveillance process.}

\begin{figure}[t]
    \centering
    \includegraphics[width=\linewidth]{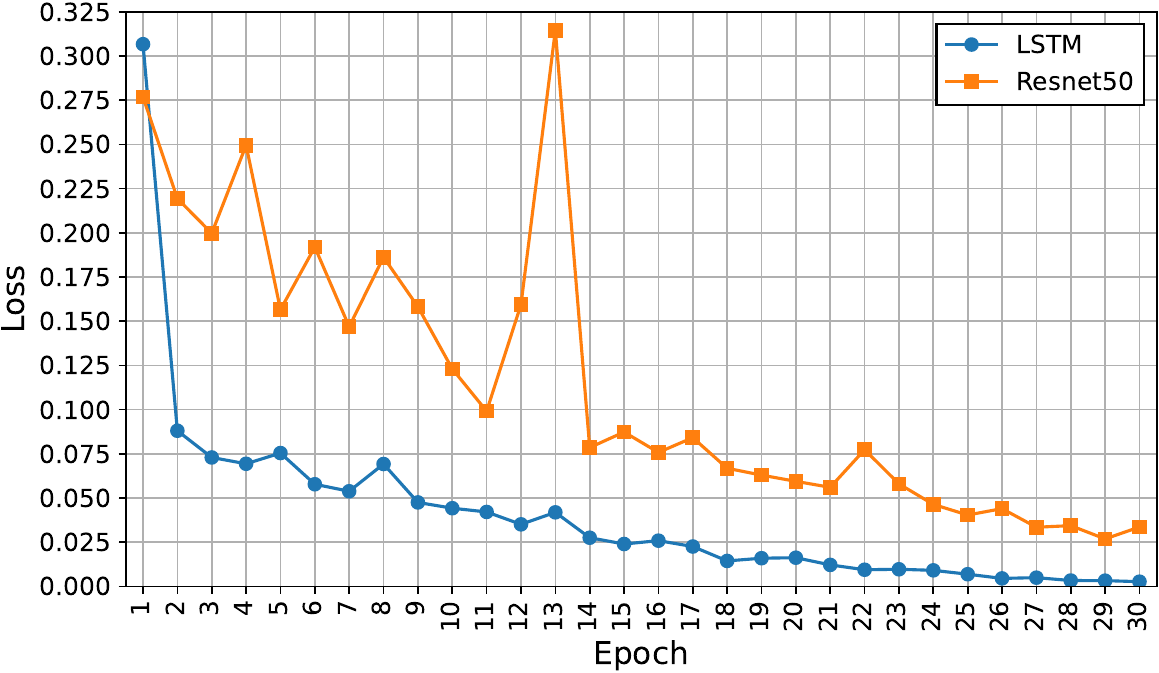}
    \caption{\textbf{Our LSTM vs. ResNet-50~\cite{liang2022hiding} loss per epoch.} Our model achieves a lower loss than the state-of-the-art.}
    \label{fig:model_loss}
    \hrulefill
\end{figure}

Recall from Section~\ref{subsec:methodology_reverse_engineering_model}, Liang et al.~\cite{liang2022hiding} model uses only ResNet-50, while our model uses both ResNet-50 and LSTM. We extracted the ResNet-50 from our model (that we used for proof-of-concept) to use it as a starting checkpoint for Liang et al.~\cite{liang2022hiding} model, so we can compare our approach with the state-of-the-art. To reproduce their model using our dataset, we used the last frame of each video as training data since it contains the position of the nozzle at that moment. \new{Our model converges faster than ResNet-50 as it reaches a lower loss over a period of 30 epochs. See ~\autoref{fig:model_loss}.  
We provide results on the instructions overhead and Z-axis normalization below.}

\begin{figure}
    \centering
    \includegraphics[width=\linewidth]{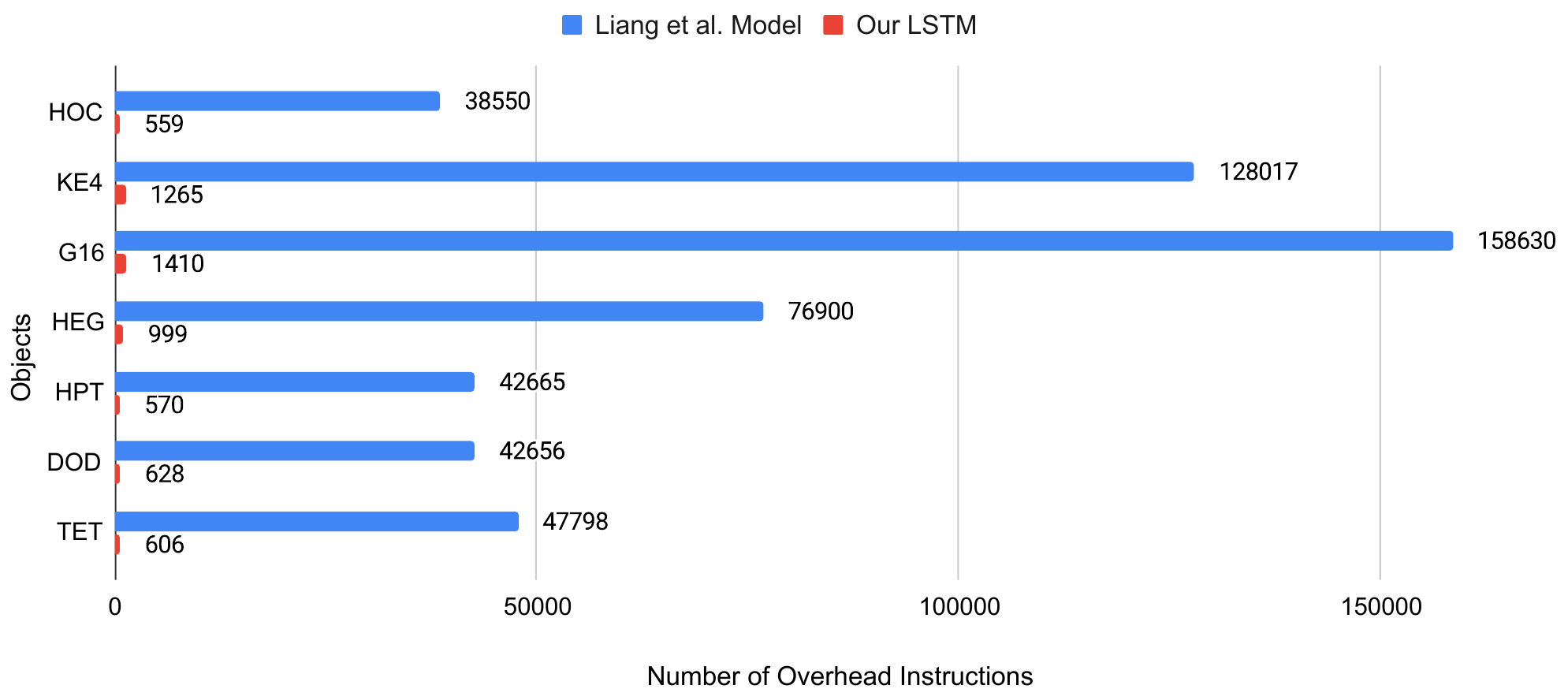}
    \caption{\textbf{Instructions overhead, as compared to the original G-code, of the G-codes recovered by our model and Liang et al. model~\cite{liang2022hiding}.} Lower instructions overhead means the recovered G-code is closer to the original in terms of the number of instructions. On average, our model produces G-codes with $30.20$ times fewer instructions than state-of-the-art~\cite{liang2022hiding}. We list object acronyms in~\autoref{tab:gec-objects}.}
    \label{fig:test-objects-size}
    \hrulefill
\end{figure}

\heading{Compact G-code.} 
Recall that our model maps a chunk of a video describing a single movement to a single point \new{on} the build plate, and \new{then} uses a sliding window technique (Section~\ref{subsec:methodology_reverse_engineering_model}) to average out the error in prediction.   Liang et al.~\cite{liang2022hiding} on the other hand map individual frames in the video to a point on the build plate. Since the number of frames in a video \new{can} be very large, the overhead in terms of the extra instructions can also be very high as shown  in~\autoref{fig:test-objects-size}. Additionally, the independent estimation of the printer's extruder location based on individual frames in the video results in a jagged estimated trajectory due to the noise in those independent estimations. However, in our model, because \new{the location estimation is based on 30 subsequent frames, the amount of noise is substantially lower resulting in a smoother more realistic trajectory.} On average, our model produces \textit{30} times fewer G-code instructions than the past work~\cite{liang2022hiding}, reducing the noise generated by predictions and producing a G-code file much more similar to the real one.

\begin{figure}[t]
    \centering
    \includegraphics[width=\linewidth]{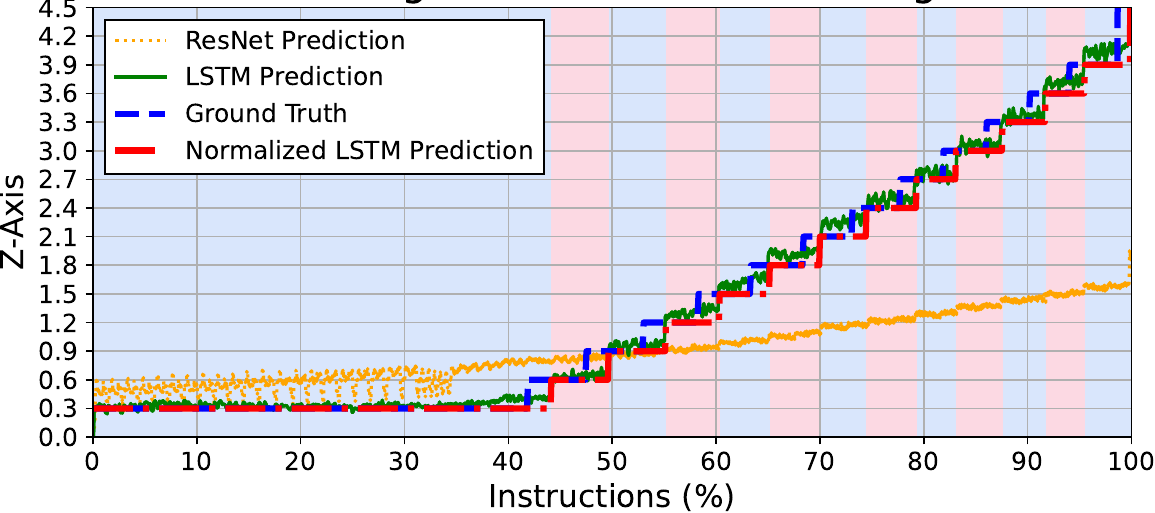}
    \caption{\textbf{Z-Axis Normalization Using PELT Change Point Detection Algorithm in G-code \new{of HPT}.} The LSTM normalized prediction (red) is very close to the ground truth (blue). While both LSTM and ResNet-50~\cite{liang2022hiding} produce noises, the former can be fixed using PELT, which does not work with the latter.}
    \label{fig:plot_z}
    \hrulefill
\end{figure}

\heading{Z-axis normalization.} Most G-codes have the z-coordinate value fixed for each layer because 3D printing is an additive manufacturing process. Therefore, we used a post-processor to normalize the z-coordinate values using the PELT change point detection algorithm. Liang et al.~\cite{liang2022hiding} mapped each frame of the video to a 3D point in the trajectory of the nozzle model and did not normalize the Z-axis. See  ~\autoref{fig:plot_z}. In blue, we show the z-coordinate values of the ground truth, i.e., z-coordinate values that are extracted from the G-code of a Hexagonal Pyramid. Observe that our LSTM prediction after normalization using the PELT change point detection algorithm almost overlaps with the ground truth. On the other hand, the raw predictions of both models have noisy values. While our raw LSTM predictions are still quite close to the ground truth, the model in ~\cite{liang2022hiding} fails to accurately recover the z-coordinates of the 3D points that describe the trajectory of the nozzle. 

\label{Comp}

\input{Sources/Sections/GEC_Eval}

%% file: Sources/Sections/GEC_Eval.tex
\subsection{G-code Equivalence Checker} 
\label{GEC}

We developed the G-code Equivalence Checker, also known as the curve checker, as a white-box solution to assess the performance of our G-code Reverse Engineering model. To that extent, we evaluated our curve checker by taking into account the core idea of the reverse-engineering process, which is to learn the mapping between chunks of videos and the coordinates of the discrete points that define the trajectory of the nozzle in each layer. That means we will check how well our curve checker compares similar and dissimilar trajectories.  

To ensure that the difference in the ground-truth trajectory (GT) and reverse-engineered trajectory (RT) is not merely due to RT being a rotated or translated version of GT, which can happen due to varying positions of the camera that is recording the print process, we need to check the rotational and translational invariance properties of our curve checker. To evaluate this, we use Datasets R, T, and RT, as described in \autoref{dtsts}, which are generated using our G-code manipulator. Since the past work~\cite{liang2022hiding} used MSE for evaluating the accuracy of their model, we also computed the normalized MSE~(nMSE)\footnote{We used the normalized version of MSE to maintain parity (in terms of scale).} between the GT and its rotated and translated versions.

\begin{figure}[!htbp]
    \centering
    \includegraphics[width=0.495\textwidth]{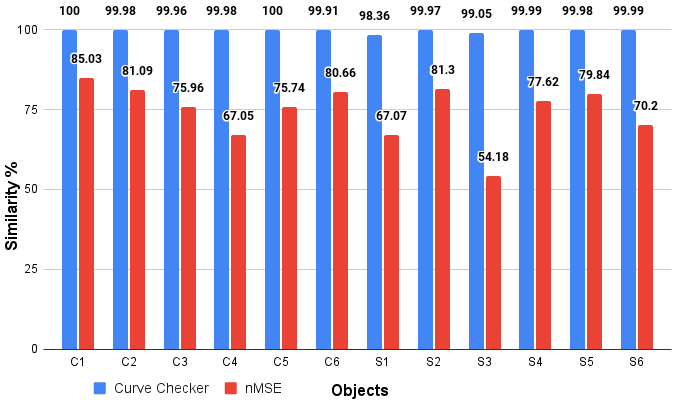}
    \caption{Rotational invariance of our G-code Equivalence checker v/s a normalized MSE-based equivalence checker}
    \label{PureRot}
    \hrulefill
\end{figure}

We begin our evaluation with Dataset R, which takes the G-code of each of the six simple and six complex objects that are placed at the center of the printing plate as the GTs and the G-codes of the object rotated 180 degrees around its centroid, at an interval of 5 degrees, as the RTs. Since each RT used here is a rotated version of the original, the similarity \% between the GTs and the RTs should be close to 100\%. 
 \autoref{PureRot} shows the performance of our curve checker and nMSE.  Across all the 12 objects, our curve checker outputs an average similarity of $98\%$ or more, where the average is taken over the layers of each object; when we take the average (of the average similarity) over all the 12 objects, we get a $99.76\%$ accuracy with a negligible standard deviation of $0.51$. This shows that our curve checker is \textit{rotationally invariant}. On the other hand, nMSE has an accuracy of $74.65\%$ with a standard deviation of $8.64$; maximum average similarity is $85.03\%$ and minimum average similarity is $54.18\%$.

Next, we use Dataset T to show that our curve checker is \textit{translationally invariant}, too. As described in \autoref{dtsts}, the recovered trajectory in Dataset T consists of G-codes of the original object translated to 36 unique points on the printing plate. A printing plate is comparable to a standard graph, with the center of the plate being equivalent to the origin of a graph. We observed that the maximum distance the largest object in our dataset could be translated, without going out of bounds, was $10\sqrt{2}$. Thus, we created a square grid, with a spacing of 4 units, bound by the lines X=10, -10 and Y=10, -10. Every lattice point on this square grid was a point of translation for this dataset. 

\begin{figure}[!htbp]
    \centering
    \includegraphics[width=0.99\linewidth]{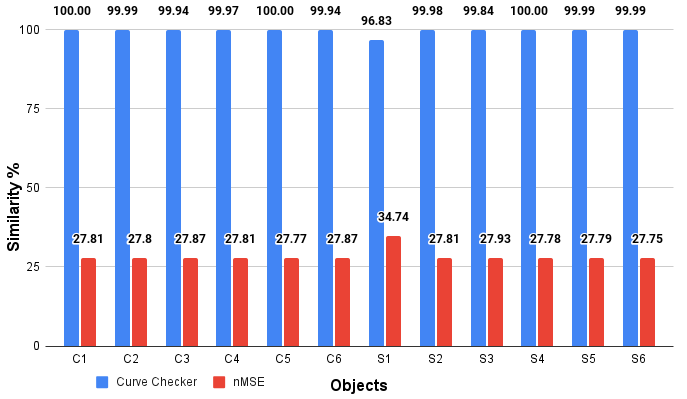}
    \caption{Translation invariance of our G-code Equivalence checker v/s a normalized MSE-based equivalence checker}
    \label{TR}
    \hrulefill
\end{figure}
\autoref{TR} shows the performance of our curve checker and nMSE in evaluating the similarity between the recovered trajectory in Dataset T and the ground truth, which is the G-code of the object placed at the origin, without rotation. Across all the 12 objects, our curve checker outputs an average similarity of $96.83\%$ or more, where the average is taken over the layers of each object; when we take the average (of the average similarity) over all the 12 objects, we get a $99.71\%$ accuracy. This shows that our curve checker is \textit{translationally invariant}. On the other hand, nMSE has a very small accuracy of $28.39\%$; maximum average similarity is $85.03\%$, and minimum average similarity is $54.18\%$.

\begin{figure}[!htbp]
    \centering
    \includegraphics[width=0.495\textwidth]{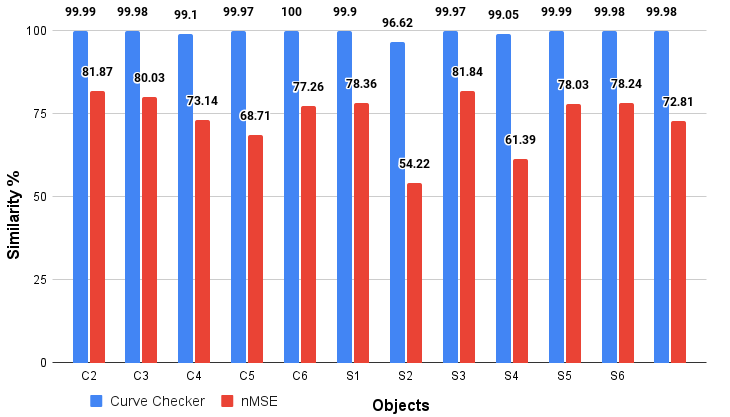}
    \caption{Translational and Rotational invariance of our G-code Equivalence checker v/s a normalized MSE-based equivalence checker}
    \label{TRot}
    \hrulefill
\end{figure}

Lastly, we show that our curve checker does not penalize models for outputting G-codes that are simultaneously rotated and translated versions of the target object. We used Dataset RT for this evaluation. In Dataset RT, we have taken 8 translation points on the printing plate: four equidistant coordinate points from the center which represent the maximum distance the largest object in the dataset can be translated, and four equidistant coordinate points from the center, which represent the minimum translation for the trajectory to show a non-negligible change. At each of these eight points of translation, the object was rotated 180 degrees, at a 5-degree interval, forming our RT Dataset. \autoref{fs1} gives a visual representation of a rotated and translated curve when compared against the ground truth.

\autoref{TRot} shows the performance of our curve checker and nMSE in evaluating the similarity between the recovered trajectory in the data set RT and the Ground Truth. Our curve checker's accuracy is  $99.54\%$, with a standard deviation of $0.98$. In contrast, nMSE has an accuracy of $73.83\%$, with a standard deviation of $8.55\%$.

%% file: Sources/Sections/discussion.tex
\section{Concluding Discussion} 
\label{Section: Conclusion}
\heading{Summary.} In the modern multi-billion dollar 3D printing industry, a 3D printable design of a low-cost and lightweight part (e.g., prosthetic) in different sectors is the IP of its manufacturer and is an attractive target for an adversary. 3D printing setups are often remotely monitored using cameras which can potentially be hacked. In this work, we gave the first optical side-channel attack which results in a complete end-to-end recovery of 3D print instructions that can be used to produce counterfeit parts. Our attack uses a machine learning model to accurately  map the video of the print process to the 3D trajectory of the printer's nozzle and uses a post-processing module to add information about extrusion and speed of the nozzle to generate printable instructions for the 3D printer. In order to evaluate the efficacy of our attack, we built a large dataset as well as a 3D printer-specific equivalence checker. We used the dataset and the equivalence checker to show that our attack has an average accuracy of 90.38\%. Additionally, we also used our attack to demonstrate the production of a counterfeit key of a master padlock \new{as well as a counterfeit 16-tooth gear in a three-gear system.}

\heading{Countermeasures.} Our reverse-engineering solution relies on the availability of the video recording of the 3D print process. We assume that an adversary can exploit known vulnerabilities in internet-facing cameras to carry out such an  attack. Naturally, the first step towards stopping such attacks will be to fix the known vulnerabilities in these cameras and implement a stringent access control policy using firewalls. Literature has suggested degrading the optical environment as a countermeasure against optical side-channel attacks. However, that will defeat the very reason why cameras were present in the first place---~ to enable remote monitoring. A potential countermeasure could be to build a verification software that works in tandem with the 3D Printer, removing the need for remote monitoring.

\heading{Reverse-engineering for remote monitoring of trojan attacks.} In a different threat model, where the 3D printer is  compromised and is used to insert ``trojans''~\cite{belikovetsky2017dr0wned} in manufactured objects, the reverse-engineering solution can be used as a defense. The operator of a 3D printer can run our solution to reverse-engineer  the 3D print instructions from a live video feed and use our equivalence checker to check if there is any significant deviation in the trajectory of the nozzle. However, in order to use our reverse-engineering solution for remotely monitoring trojan attacks on real-world complex objects, it will be necessary to first evaluate our model's performance on complex objects and design more sophisticated models if needed. Another avenue of improving the reverse-engineering solutions is to use multiple sources of side-channel information~\cite{gao2018watching}. Note that advances in these techniques will also help an IP-stealing adversary. Hence, the manufacturers should exercise caution while deploying these solutions.

\new{\heading{Limitations.} Constructing a dataset of 3D-print videos paired with ground-truth G-code is a time-intensive process. This bottleneck makes it difficult to explore more powerful architectures such as transformers~\cite{vivit-vision-transformers}.  However, transformers are known to be data-hungry and prone to overfitting when trained on small datasets, which can lead to poor generalization. Another limitation lies in the curve checker metric, which, while effective at measuring trajectory similarity in a rotation-and-translation-invariant way, lacks interpretability with respect to structural implications. Specifically, it does not pinpoint whether the discrepancies arise from dimensional inaccuracies, infill deviations, or layer height differences. This gap in diagnostic resolution is noteworthy because validating structural integrity typically requires expensive and destructive mechanical tests such as fatigue analysis. If the trajectory-level mismatches could be used to predict high-level structural faults, then costly physical testing could be partially avoided or better targeted, improving the overall efficiency of reverse-engineering analysis. Moreover, our curve checker uses a convex hull (under the hood) that can have approximation errors which in turn can result in incorrect distance values between the ground truth and the reverse-engineered G-codes. While we validated our results using visual proofs via plotting both trajectories, future works can explore alternative algorithms such as concave hull~\cite{moreira2007concave}. } 

\heading{Ethical considerations.} In our experiments, we did not use any 3D model design that was copyrighted or belonged to a protected IP category. We obtained the 3D model of all objects used in the evaluation from open-source platforms, namely Thingiverse~\cite{thingiverse} and GrabCAD~\cite{grabcad}. For 3D printing, we use default settings and configurations of the 3D printer (Geeetech A20T) as well as the slicer software (Ultimaker Cura). For our experiments on the padlock key, as shown in Subsection \ref{subsec:reverse_engineering_model_eval}, we purchased a lock and key set. Then, we created a custom G-code design of that key and treated that as our ground truth. Furthermore, all G-code parameters chosen were done based on our understanding and exploration of slicers and slicer manuals and do not include any private or confidential data.

%% file: Sources/Sections/acknowledgement.tex
\begin{acks}
The authors acknowledge funding from AI Manufacturing Pilot Facility project under Georgia Artificial Intelligence in Manufacturing (Georgia AIM) from the Economic Development Administration, Award 04-79-07808. Co-Author Marco Garza was supported by the U.S. Department of Energy/National Nuclear Security Administration (DOE/NNSA) \#DE-NA0003985.
Any opinions, findings, conclusions, or recommendations expressed in this material are those of the authors and do not necessarily reflect the views of the funding agency.

\end{acks}